\documentclass[aps,prb,preprint,showpacs]{revtex4-1}
\usepackage{amsmath}
\usepackage{amssymb}
\usepackage{bm}
\usepackage{epsfig}
\usepackage{graphicx}
\usepackage{color}

\usepackage{color}
\usepackage{colortbl}

\def\be{\begin{equation}}
	\def\ee{\end{equation}}
\def\bea{\begin{eqnarray}}
	\def\eea{\end{eqnarray}}

\begin{document}
	\title{Polarized emission of CdSe nanocrystals in magnetic field: the role of phonon-assisted recombination of the dark exciton}
	
	\author{Gang Qiang,\textit{$^{1}$} Aleksandr A. Golovatenko,\textit{$^{2}$}  Elena V. Shornikova,\textit{$^{1}$} Dmitri R. Yakovlev,\textit{$^{1,2}$} Anna V. Rodina,\textit{$^{2}$} Evgeny A. Zhukov,\textit{$^{2}$} Ina V. Kalitukha, Victor F. Sapega,\textit{$^{2}$} Vadim K. Kaibyshev,\textit{$^{2}$}  Mikhail A. Prosnikov,\textit{$^{3}$}  Peter C. M. Christianen,\textit{$^{3}$} Aleksei A. Onushchenko,\textit{$^{4}$} and Manfred Bayer\textit{$^{1,2}$}}
	
	\affiliation{
		$^1$Experimentelle Physik 2, Technische Universit\"at Dortmund,~44227 Dortmund, Germany \\
		$^2$Ioffe Institute, Russian Academy of Sciences, 194021 St. Petersburg, Russia \\
		$^3$High Field Magnet Laboratory (HFML-EMFL), Radboud University, 6525 ED Nijmegen, The Netherlands \\
		$^4$ITMO University, 199034 St. Petersburg, Russia
	}

	\date{\today}

 \begin{abstract}
 	 The recombination dynamics and spin polarization of excitons in CdSe nanocrystals synthesized in a glass matrix are investigated using polarized photoluminescence in high magnetic fields up to 30 Tesla. The dynamics are accelerated by increasing temperature and magnetic field, confirming the dark exciton nature of the low-temperature photoluminescence (PL). The  circularly polarized PL in magnetic fields reveals several unusual appearances: (i)  a spectral dependence  of the polarization degree, (ii) its low saturation value, and (iii) a stronger intensity of the Zeeman component which is higher in energy. The latter feature is the most surprising being in contradiction with the thermal population of the exciton spin sublevels. The same contradiction was previously observed in the ensemble of wet-chemically synthesized CdSe nanocrystals, but was not understood. We present a theory which explains all the observed features and shows that the inverted ordering of the circular polarized PL maxima from the ensemble of nanocrystals is a result of competition between the zero phonon (ZPL) and one optical phonon (1PL) assisted emission of the dark excitons. The essential aspects of the theoretical model are different polarization properties of the dark exciton emission via ZPL and 1PL recombination channels  and the inhomogeneous broadening of the PL spectrum from the ensemble of nanocrystals exceeding the optical phonon energy.

 \end{abstract}

\maketitle

\section{Introduction}
Since a few decades colloidal semiconductor nanocrystals (NCs) are in the focus of intensive research. Due to the continuous progress in technology, nanocrystals with different sizes, shapes, compositions, surface properties have been synthesized.\cite{Ekimov1981,Murray1993,MiCiC1994,XGPeng2000,Kovalenko2015,Ithurria2008,Lesnyak2013,Fedin2016}  
Understanding of their optical, electrical and chemical properties has led to applications in various fields, such as light-emitting diodes, \cite{Colvin1994, Kim2011}  laser technology, \cite{Klimov2000} field-effect transistors, \cite{Talapin2005}  solar cells, \cite{Carey2015, Pietryga2016} biological labels,  \cite{Marcel1998,Efros2018} etc. 

For direct bandgap semiconductor NCs, e.g. made of CdSe, CdTe or InP,  most of the optical properties can be explained within the exciton fine structure model. \cite{Efros1996,Efros2003, Goupalov, Sercel}  In ideal spherical NCs the band-edge exciton state $1S_{3/2}1S_e$ is eight-fold degenerated. $1S_{3/2}$ is the lowest quantum size level of the holes, which is four-fold degenerate with respect to its total angular momentum components. $1S_e$ is the lowest quantum size level of the electrons, which is doubly degenerate with respect to its spin components. However, due to the NC shape asymmetry, the intrinsic crystal field, and the electron-hole exchange interaction, the degeneracy is lifted resulting in the formation of five exciton states. In the case of wurtzite CdSe NCs, the two lowest states are typically optically-forbidden (dark) exciton state $\left| F\right\rangle$ (angular momentum projection along the NC quantization axis $\pm2$) and a higher-lying optically-allowed (bright) exciton state $\left|A\right\rangle$ (momentum projection $\pm1$). \cite{Efros2003} The presence of the bright and dark excitons allows one to explain the recombination dynamics, spin dynamics, and  magneto-optical properties, taking into account also the Zeeman effect.  \cite{Halperin2001,Furis2005,Labeau2003,Liu2014,Efros1996,Biadala2010,Brodu2019,Brodu2019_1}

CdSe NCs are a test bed for investigation of colloidal NCs, therefore,  their optical properties have been extensively studied. In particular, the following magneto-optical properties in external magnetic fields of wet-chemically synthesized colloidal CdSe NCs were reported: shortening of the dark exciton lifetime,\cite{Nirmal1995,Efros1996} circular polarization of photoluminecsence (PL),\cite{Halperin2001,Furis2005,Wijnen2008,Granadosdelguila2017} exciton fine structure splitting including the Zeeman effect in single NCs,\cite{Biadala2010,Fernee2014} anisotropic exchange interaction,\cite{Furis2006,Htoon2009} and electron spin coherence.\cite{Gupta2002,Hu2019} No magneto-optical study has been performed so far on CdSe in glass matrix.    

Studies of the circularly polarized PL of CdSe-based NCs in magnetic field~\cite{Furis2005, Wijnen2008} revealed a puzzling behavior: with increasing field the ${\sigma}^{-}$ polarized PL, which is stronger in intensity compared to the ${\sigma}^{+}$ component, shifts to higher energy than the ${\sigma}^{+}$ polarized emission. This contradicts with the expectation for the thermal population of the exciton spin sublevels split in the magnetic field. Later, the same contradiction was reported for colloidal CdTe NCs.\cite{Liu2014} This contradiction, however, is absent in fluorescence line narrowing (FLN) experiments and in measurements of single NCs, i.e. for the experimental conditions when the inhomogeneous broadening of the PL due to NC size dispersion in the ensemble is suppressed.~\cite{Wijnen2008,Biadala2010} In the FLN experiments, pronounced optical phonon assisted recombination of the dark exciton was observed.\cite{Nirmal1994,Nirmal1995,Norris1996,Woggon1996,Efros1996,Wijnen2008} Under nonresonant excitation the low-temperature PL of the NC ensemble is contributed by zero phonon and optical phonon-assisted exciton emission, which form overlapping bands due to NCs of different size.\cite{Norris1996,Efros1996} In Refs.~\onlinecite{Furis2005,Wijnen2008} the contradiction with respect to the energy shift was tentatively assigned to the interplay between different emission channels of the dark exciton. However, a corresponding theoretical model has not been proposed so far.

For the present magneto-optical study, we choose CdSe NCs which are synthesized in glass matrix. One of the motivations is to compare their magneto-optical properties with the wet-chemically synthesized CdSe NCs. Further reasons are based on the simpler experimental situation, namely the simpler surface conditions due to the absence of organic ligand passivation, long-term stability due to the matrix encapsulation, and the ideally random orientation of the NC quantization axis in the glass-based NC ensemble which allow us to make an easier model description.

In this paper, we study CdSe NCs in glass with diameters varying from 3.3 nm up to 6.1 nm and  investigate the recombination dynamics and spin polarization properties of excitons in strong magnetic fields up to 30~T. We find experimentally the same contradiction between intensities and energies of the circularly polarized PL as reported for wet-chemically prepared CdSe NCs, and a strong spectral dependence of the circular polarization degree. We develop a theoretical model accounting for the contributions of both zero phonon emission (ZPL) and first optical phonon-assisted emission (1PL) of the dark excitons, which allows us to describe all experimental signatures.

%\subsection{This is the subsection heading style}
%Section headings can be typeset with and without numbers.\cite{Abernethy2003}

%\subsubsection{This is the subsubsection style.~~} These headings should end in a full point.  

%\paragraph{This is the next level heading.~~} For this level please use \texttt{\textbackslash paragraph}. These headings should also end in a full point.

\section{Experimental results and modeling}
\subsection{Time-integrated and time-resolved photoluminescence} 
\begin{figure*}[h!]
	\centering
	\includegraphics[width= 17.2 cm]{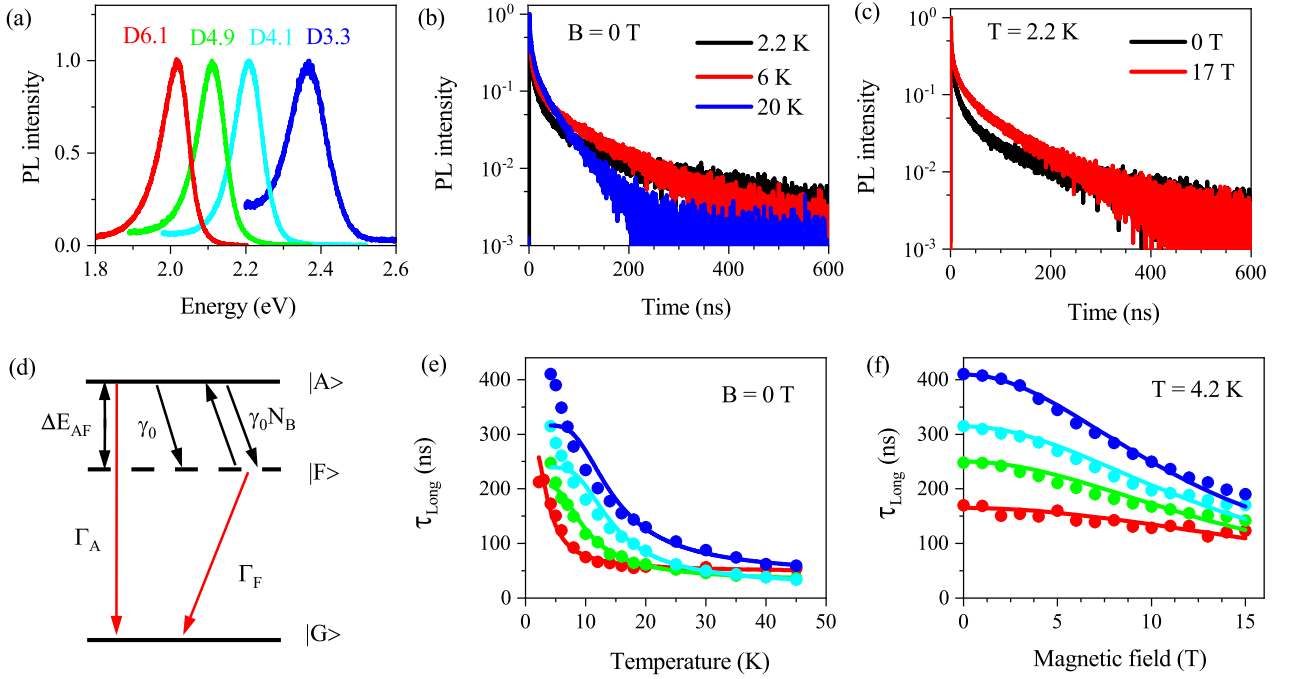}
	\caption{Photoluminescence and recombination dynamics. (a) PL spectra of the studied CdSe NCs measured at $T=4.2$~K. (b) Recombination dynamics measured for  the sample D6.1 at various temperatures, $B=0$~T. (c) Recombination dynamics measured for the sample D6.1 at $B=0$ and 17~T, $T=2.2$~K. (d) Scheme of exciton energy levels. $|A\rangle$, $|F\rangle$, and $|G\rangle$ denote bright and dark exciton states, and crystal ground state, respectively. ${\Delta}E_{AF}$ is the bright-dark splitting. ${\gamma}_{0}$ is the zero-temperature relaxation rate from the bright to the dark exciton state. ${{\gamma}_{0}}{N_{B}}$ is the thermal activation  rate of the reverse process. ${\Gamma}_{A}$  and ${\Gamma}_{F}$ are the recombination rates of the bright and dark excitons. (e) Temperature dependence of the long component of the PL decay, $\tau_{Long}$, for all samples at $B=0$~T. The lines are fits with eq~(S5$^\dag$). (f) Magnetic field dependence of $\tau_{Long}$ at $T=4.2$~K for all samples. The lines show the averaged lifetime of the dark exciton calculated with eq~(S23$^\dag$) in a randomly oriented ensemble of NCs as described in  ESI Section~S4.2$^\dag$. The color codes in panels (e,f) are same as in panel (a).}  
	\label{fig:fig1}
\end{figure*}

We studied four samples with CdSe NCs synthesized in a silicate glass matrix (see Methods). The sample parameters are given in Table~\ref{tab:table1}. Their photoluminescence (PL) spectra measured at $T=4.2$~K are shown in Figure~\ref{fig:fig1}a. The quantum confinement enhancement with reducing NC diameter shifts the emission band to higher energies from 2.017~eV for D6.1 up to 2.366~eV for D3.3. The full width at half maximum (FWHM) of the PL spectra is about 100~meV, which is caused by the dispersion of NC sizes. Absorption spectra are shown and compared with the PL in Figure~S1$^\dag$. The Stokes shift between the exciton resonance in absorption and PL maximum is also size-dependent, it is larger in smaller NCs, e.g. 89~meV in D3.3 and only 22~meV in D6.1.

\begin{table*}[h!]
	\centering
	\small
	\caption {\ Parameters of CdSe NCs in glass measured at $T=4.2$~K. }
	\begin{tabular*}{1\textwidth}{@{\extracolsep{\fill}}lllll}
		\hline
		Sample & D3.3 & D4.1 & D4.9 & D6.1 \\
		\hline
		NC diameter (nm) & $3.30\pm0.17$ & $4.10\pm0.21$ & $4.9\pm0.25$ & $6.1\pm0.31$ \\
		PL FWHM (meV) &  120  & 99 & 96 & 101 \\
		PL peak energy  (eV) &  2.366  & 2.210 & 2.111 & 2.017 \\
		Absorption peak energy (eV) &  2.455  & 2.267 & 2.160 & 2.039 \\
		Stokes shift  (meV) &  89  & 57 & 49 & 22 \\	    
		Bright-dark splitting from PL decay  (meV) &  3.9  & 4.3 & 2.6 & 0.9\\	
		Bright-dark splitting from FLN  (meV) &-&8.4&7.5&3.5\\
		\hline
	\end{tabular*}
	\label{tab:table1}
\end{table*}

The low-temperature PL in the studied samples is dominated by exciton recombination, which has a characteristic decay variation with temperature and magnetic field. This allows one to distinguish neutral and charged excitons in the NCs.\cite{Liu2013,Shornikova2018,Shornikova2020NL} Taking sample D6.1 as a representative example, we show in Figure~\ref{fig:fig1}b the recombination dynamics at various temperatures. With increasing temperature from 2.2 to 20~K, the decay time of the long component, $\tau_{Long}$, shortens from 212~ns down to 58~ns. The temperature dependence of $\tau_{Long}$ for all studied samples is given in Figure~\ref{fig:fig1}e. One can see that the decay times shorten with increasing temperature as a result of thermal activation from the dark exciton to the bright exciton state. A similar behavior is observed when external magnetic field is applied, in Figure~\ref{fig:fig1}c the long component measured at $T=2.2$~K shortens from 212~ns down to 119~ns with increasing field up to $B=17$~T. The magnetic field dependence of $\tau_{Long}$ for all studied samples is given in Figure~\ref{fig:fig1}f. 

This is the typical behavior for exciton emission in CdSe NCs and can be well explained with the diagram in Figure~\ref{fig:fig1}d (a detailed scheme of the exciton levels is given in Figure~S13a$^\dag$). The optically-forbidden dark state $|F\rangle$ with angular momentum $\pm2$ is the exciton ground state. The closest optically-allowed bright state $|A\rangle$ with $\pm1$ is shifted to higher energy by the bright-dark splitting ${\Delta}E_{AF}$.\cite{Efros1996} The fast component of the PL decay is provided by the recombination of bright excitons with rate $\Gamma_{A}$ and their rapid thermalization to the dark state with rate $\gamma_{0}(1+N_B)$, where $\gamma_{0}$ is the spin-flip rate, ${N_B} =1/[\exp(\Delta E_{AF}/{k_B}T)-1]$ is the Bose-Einstein phonon occupation at temperature $T$. The long component corresponds to the dark exciton recombination with rate $\Gamma_{F}$. With increasing temperature, the dark excitons are activated to the bright state. In the case of a one-phonon process, the activation rate is given by $\gamma_{0}{N_B}$. The shortening of the long component in a magnetic field is due to the mixing of bright and dark exciton states by the field component perpendicular to NC quantization axis.\cite{Efros1996,Liu2013} These properties allow us to uniquely assign the long decay component $\tau_{Long}$ to the recombination of the dark excitons. More information about the PL decay fitting is given in ESI Section~S2$^\dag$.

The analysis of the $\tau_{Long}$ temperature dependence allows us to evaluate the bright-dark splitting ${\Delta}E_{AF}$ (see ESI Section~S2$^\dag$). The corresponding fits are shown by solid lines in Figure~\ref{fig:fig1}e, and the evaluated ${\Delta}E_{AF}$ are given in Table~\ref{tab:table1}. In Figure~S6c$^\dag$, we compare the ${\Delta}E_{AF}$ with the literature data\cite{Norris1996,Woggon1996,Chamarro1996,Efros1996} and a good consistency is found. However, ${\Delta}E_{AF}$ in D3.3 is deviating from the general trend, which may point toward a slightly prolate shape of the CdSe NCs.\cite{Rodina2018FTT} The $\Delta E_{AF}$ values determined from fluorescence line narrowing  (see ESI Section~S2.3$^\dag$) are found to be larger as compared to the values determined from the $\tau_{Long}$ temperature dependence. Two explanations for this discrepancy were proposed: (i) The $\Delta E_{AF}$ determined from the $\tau_{Long}$ temperature dependence correspond to the true values of the bright-dark splitting, while the $\Delta E_{AF}$ from FLN can show an additional energy shift due to polaron formation\cite{Biadala2017}, (ii) The $\Delta E_{AF}$ determined from the $\tau_{Long}$ temperature dependence correspond to the energy of confined acoustic phonons, which provide thermal activation to the bright excitons.\cite{Oron2009}  

\begin{figure*}[h!]
	\centering
	\includegraphics[width= 17.2 cm]{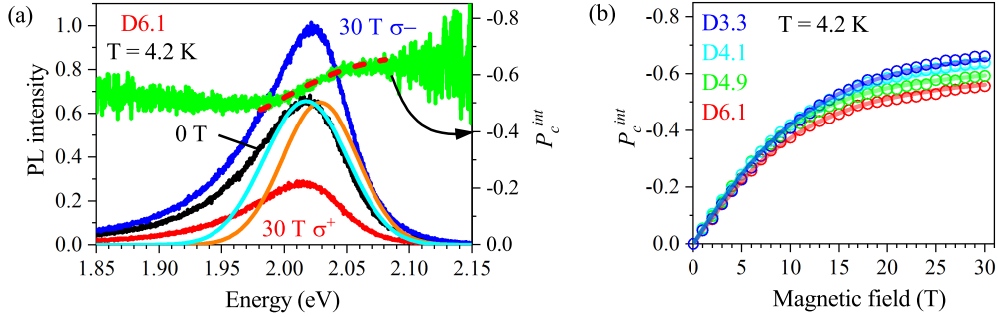}
	\caption{Polarized photoluminescence in magnetic field. (a) ${\sigma}^{-}$ (blue) and ${\sigma}^{+}$ (red) circularly polarized emission, measured at $B=30$~T for the sample D6.1.  The black spectrum at $B=0$~T is shown for comparison. Green and red dashed curves show the experimental and calculated spectral dependence of the time-integrated DCP, respectively. The orange curve is the function $f_{ZPL}(E)$ and the cyan curve is the calculated PL spectrum at $B=0$~T, which accounts for the ZPL and 1PL contributions. (b) Magnetic field dependences of $P_c^{int}(B)$ measured at the PL maximum in all samples. Lines are fits with equations \eqref{eq:intpm} and \eqref{eq:pcint}.}
	\label{fig:fig2}
\end{figure*}

\subsection{Polarized exciton emission in strong magnetic fields} Let us turn to the main focus of this paper, i.e. the polarized exciton emission in strong magnetic fields. Figure~\ref{fig:fig2}a shows circularly-polarized PL spectra of the sample D6.1 measured at $T=4.2$~K and $B=30$~T. One can see that compared to the black spectrum at $B=0$~T, the intensity of the ${\sigma}^{-}$ component (blue) is increased while it is decreased for the ${\sigma}^{+}$ component (red). Figure~\ref{fig:fig2}b shows the magnetic field dependence of the time-integrated degree of circular polarization (DCP), $P_{c}^{int}$, calculated with eq~\eqref{eq:pc}, measured at the PL maximum. In the sample D6.1 $P_{c}^{int}(B)$ increases linearly in low magnetic fields and saturates in high fields at ${P_c}^{int}= -0.56$.

The green line in Figure~\ref{fig:fig2}a shows the spectral dependence of ${P_c}^{int}$, which is larger at the high energy side ($-0.63$), decreases around the PL peak energy, and keeps nearly constant ($-0.53$) at the low energy side. This is quite unusual as one does not expect to have a considerable spectral dependence of the dark exciton $g$-factor, $g_F$, within the emission line. In fact, as we will show below (Table~S3$^\dag$), $g_F$  is about constant for all studied NCs covering a much larger spectral range. It is also expected that in the ensemble of NCs with randomly oriented c-axis (the case for NCs in glass), the DCP saturation value should reach $-0.75$. However, as one can see in Figure~\ref{fig:fig2}b, this is obviously not the case, ${P_c}^{int}$ does not reach $-0.75$ and has a clear size-dependence. At $B=30$~T, ${P_c}^{int}= -0.56$ in D6.1 and $-0.66$ in D3.3. 

\begin{figure*}[h!]
	\centering
	\includegraphics[width= 17.2 cm]{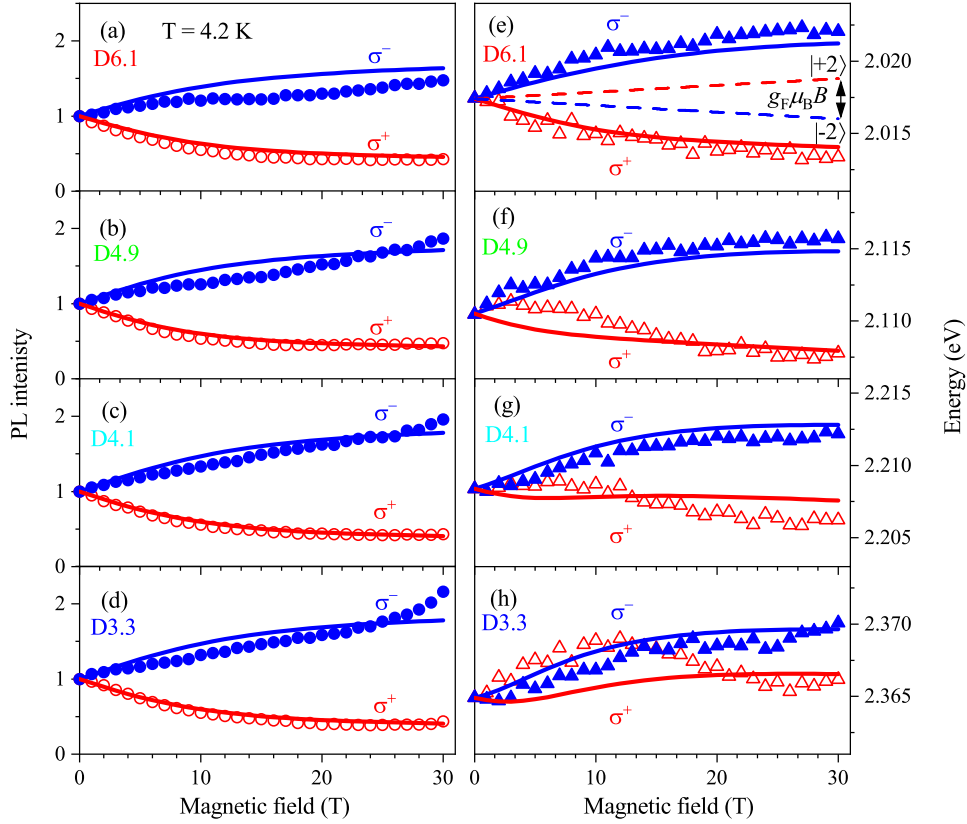}
	\caption{Photoluminescence intensity and spectral shifts in magnetic fields. (a-d) PL intensity of the ${\sigma}^{+}$ (red) and ${\sigma}^{-}$ (blue) polarized PL as function of magnetic field in the CdSe NCs. (e-h) Magnetic field dependences of the corresponding PL peak energies. For all panels, the symbols correspond to the experimental data, the lines show the modeling results. Dashed lines in panel (e) show the Zeeman splitting of the dark exciton spin sublevels $-2$ (blue) and $+2$ (red) in a nanocrystal with c-axis parallel to the magnetic field direction.}
	\label{fig:fig3}
\end{figure*}

Figures~\ref{fig:fig3}a-d show the magnetic field dependence of the polarized PL intensity. The  ${\sigma}^{-}$ polarized component increases monotonously with magnetic field and the ${\sigma}^{+}$ polarized component remains almost constant after an initial decrease. Most surprising are the spectral shifts of the ${\sigma}^{+}$ and ${\sigma}^{-}$ polarized components (Figures~\ref{fig:fig3}e-f). For neutral excitons, it is expected that the PL maximum of the stronger component, i.e. ${\sigma}^{-}$, shifts in magnetic field to lower energy, as the lower energy spin state of the exciton has the higher thermal population. This expectation is shown by the dashed lines for the sample D6.1 in Figure~\ref{fig:fig3}e (more details will be given below), but the experimental shift of the PL maximum demonstrates a very different behavior. First, for the D6.1, D4.9 and D4.1 samples, the ${\sigma}^{+}$ component shifts to lower energy and second, the splitting between ${\sigma}^{+}$ and ${\sigma}^{-}$ is several times larger than expected (Figures~\ref{fig:fig3}e-g). In the sample D3.3, both components shift initially to higher energy and the ${\sigma}^{+}$ lowers its energy for fields above 15~T (Figure~\ref{fig:fig3}h). 

\subsection{Model description} 
\begin{figure*}[h!]
	\centering
	\includegraphics[width=17.2 cm]{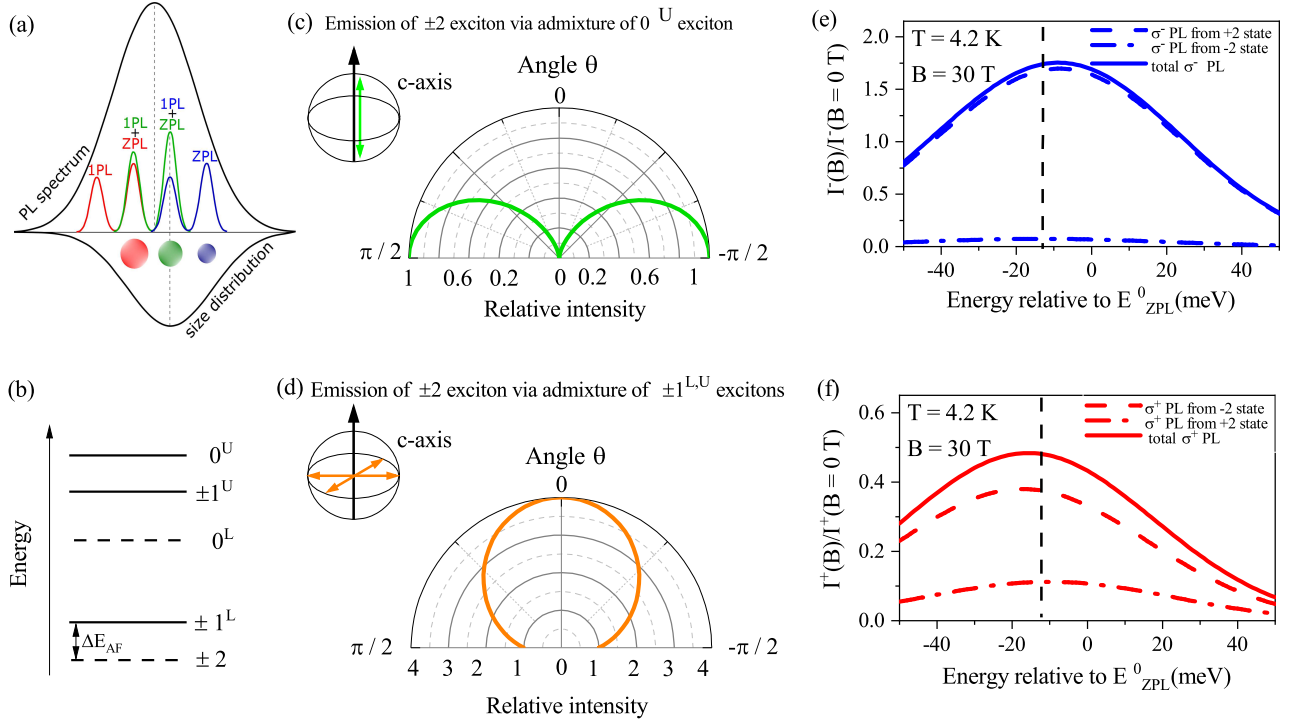}
	\caption{Concept and theory. (a) Schematics representing the ZPL and 1PL contributions in an inhomogeneous ensemble of NCs. (b) Fine structure of $1S_{3/2}1S_e$ exciton. Dashed lines show dark excitons, solid lines show bright excitons. The properties of the $\pm2$ dark exciton emission are acquired via the admixture of (c) the $0^U$ bright exciton and (d) the $\pm1^{L,U}$ bright excitons. The spatial profiles of the emission intensity are plotted as functions of the angle $\theta$ between the c-axis of a nanocrystal and the direction of light propagation. Green and orange arrows show the orientation of one-dimensional and two-dimensional dipoles with respect to the c-axis of a NC. Calculated $\sigma^-$ (e) and $\sigma^+$ (f) polarized PL spectra for sample D6.1 (solid lines) at $T=4.2$~K and $B=30$~T. Dashed (dash-dotted) lines show the contributions from the $-2$ ($+2$) dark exciton state. Vertical dashed line shows the position of the PL maximum at $B=0$~T.}
	\label{fig:cartoon}
\end{figure*}

In order to resolve the contradiction between intensities and energies of the polarized PL, we extend the theoretical model for the circular polarized emission from an ensemble of randomly oriented nanocrystals first developed in Ref.~ \onlinecite{Halperin2001}. The extension accounts for the linearly polarized contribution coming from the dark exciton recombination assisted by optical phonons. For the spectrally broad PL bands in CdSe NCs of about 100~meV width, which is four times the optical phonon energy in CdSe (26~meV in bulk), the PL is composed of the ZPL emission of the dark excitons from NCs of one size and optical phonon-assisted emission of the dark excitons from smaller NCs, as illustrated in Figure~\ref{fig:cartoon}a. The contribution of acoustic phonon-assisted recombination of the dark excitons cannot be resolved within the ZPL line, however, is can be seen by single nanocrystal spectroscopy.\cite{Ferne2008} The ZPL emission usually has properties of a two-dimensional dipole, indicating the activation of the $\pm2$ dark exciton through admixture with the $\pm1^{L,U}$ bright excitons (see Figure \ref{fig:cartoon}d).\cite{Empedocles1999} However, the specific mechanism of the ZPL recombination is still under debate.\cite{Rodina2016, Leung1998, Califano2005}  It was shown in Ref.~\onlinecite{Rodina2016} that the dark exciton recombination with the assistance of optical or acoustic phonons results in predominantly linearly polarized emission, corresponding to the admixture of the $0^U$ bright exciton (Figure~\ref{fig:cartoon}c). Thus, the ZPL and phonon-assisted emission have different spatial distribution profiles of the emission, which are determined by the relative orientation between the direction of the light propagation and direction of the anisotropic c-axis of wurtzite CdSe nanocrystals.\cite{Efros1996,Empedocles1999,Rodina2016} In the case of a randomly oriented ensemble of nanocrystals, all these factors modify the magnetic field and spectral dependences of $P_c^{int}$.

The modeling results presented in Figures~\ref{fig:fig2} and \ref{fig:fig3} are based on the consideration of the ZPL and one optical phonon assisted line (1PL) contributions to the PL. The contribution of the linearly polarized recombination of the dark exciton with assistance of acoustic phonons is considered in ESI Section~S4$^\dag$. Our task is to calculate the $\sigma^+$ and $\sigma^-$ polarized PL spectra and the spectral dependence of $P_c^{int}$. For this purpose we sum the ZPL and 1PL contributions, which have different spectral distributions, and average them over the random orientation of the hexagonal c-axis in an ensemble of NCs:
\begin{align}
	\label{eq:intpm}
	I^{\pm}(E,B)=\int_{0}^{1}dx\sum_{i=\pm 2} I_{i,ZPL}^{\pm}(x)f_{\rm ZPL}(E-\nonumber \\ -\delta E_{i}(x,B)) +I_{i,1PL}^{\pm}(x)f_{\rm 1PL}(E-\delta E_{i}(x,B)) .
\end{align}
Here $E$ is the spectral energy,  $\delta E_{\pm2}(x,B)=\pm g_F\mu_BBx/2$ are the Zeeman shifts of the $\pm2$ dark exciton states,  $x=\cos \theta$ with $\theta$  being the angle between the c-axis of the nanocrystal and magnetic field applied in the Faraday geometry. The explicit forms of $I_{i,ZPL}^{\pm}(x)$ and $I_{i,1PL}^{\pm}(x)$ are given in ESI Section~S4$^\dag$. To account for the inhomogeneous broadening of the PL spectrum due to size dispersion of the NCs, we assume the following distribution functions of the ZPL and 1PL emission:
\begin{align}
	f_{ZPL}(E) &= \frac{1}{w\sqrt{2\pi}} \exp\left( -\frac{(E-E_{ZPL}^0)^2}{2w^2}\right)	 \, , \\
	f_{1PL}(E) &= \frac{1}{w\sqrt{2\pi}} \exp\left( -\frac{(E-E_{1PL}^0)^2}{2w^2}\right)= \nonumber \\	 &=f_{ZPL}(E+E_{LO}) \nonumber .
\end{align}
Here $E_{ZPL}^0$ corresponds to the energy of the ZPL in QDs at the maximum of the size distribution, $E_{LO}=26$~meV is the energy of the optical phonon in CdSe, $w$ is the standard deviation. The function $f_{\rm ZPL}(E)$ used for fitting of the experimental data for the D6.1 sample is shown  in Figure~\ref{fig:fig2}a by the orange line. The cyan line models the PL spectrum at $B = 0$ with accounting for the 1PL contribution.     

Using eq~\eqref{eq:intpm} we calculate the DCP spectral dependence as:
\begin{align}
	\label{eq:pcint}
	P_c^{int}(E,B)=\frac{I^{+}(E,B)-I^-(E,B)}{I^{+}(E,B)+I^-(E,B)} .
\end{align}
The fit parameters are: the dark exciton $g$-factor $g_F$, the characteristic energy of interaction $\varepsilon$ which results in the admixture of the dark and  $\pm1^L$ bright exciton states in zero magnetic field, the ratio of the 1PL and ZPL recombination rates at zero magnetic field $\chi_0$, and a phenomenological parameter $c_{1PL}$, which determines the increase of the 1PL recombination rate in magnetic field, while the increase of the ZPL recombination rate in magnetic field is described within second order perturbation theory \cite{Efros1996,Rodina2016}. The parameter $\chi_0$ can be estimated independently from the ratio of the 1PL to ZPL intensities measured in fluorescence line narrowing experiments. Typically this ratio is of the order of unity.\cite{Nirmal1994,Nirmal1995,Norris1996,Woggon1996,Efros1996,Wijnen2008} The FLN spectra for samples D4.1, D4.9 and D6.1 are shown in Figure~S7$^\dag$. Fitting of the magnetic field dependence of the long decay time (Figure~\ref{fig:fig1}f) imposes a restriction on the $\chi_0$, $\varepsilon$ and $c_{1PL}$ values. Finally, the linked sets of parameters $\chi_0$, $\varepsilon$ and $c_{1PL}$ together with $g_F$ are used for joint fitting of the magnetic field dependences of $P_c^{int}$, the 
intensities and maxima positions of the $\sigma^-$ and $\sigma^+$ polarized PL.

According to the fit results, the dark exciton $g$-factor $g_F\approx1.7$ is almost independent of the nanocrystal diameter (Table~S3$^\dag$). We note that the DCP in high magnetic fields does not saturate at $-0.75$ predicted in Ref.~\onlinecite{Halperin2001} for a randomly oriented ensemble of NCs. The reason is the contribution of the predominantly linearly polarized 1PL emission of dark excitons. This contribution is found to be stronger in large nanocrystals (Table~S3$^\dag$), resulting in a smaller saturation value of the DCP (Figure~\ref{fig:fig2}b), and is consistent with the relative intensities of the 1PL and ZPL emission determined from FLN (see Figure~S7$^\dag$).

Besides the saturation of the DCP in high magnetic fields, the 1PL emission causes the DCP spectral dependence (Figure~\ref{fig:fig2}a). The reason is the weaker relative intensity of the 1PL emission at the high-energy side of the PL spectrum due to the small number of NCs that can provide this contribution. The pronounced spectral dependence of the DCP in large NCs correlates with its low saturation level, i.e. with the strong 1PL emission (Figure~S17$^\dag$).

\subsection{Modeling of photoluminescence spectra and circular polarization} We turn now to the modeling of the intensities and maxima positions of the $\sigma^+$ and $\sigma^-$ polarized PL. The calculated magnetic field dependences of the PL intensities (Figures~\ref{fig:fig3}a-d) are in a good agreement with the experimental data for all studied samples. In a randomly oriented ensemble of NCs, the emission of $\sigma^+$ polarized photons from the predominantly populated $-2$ dark exciton state results in a saturation of the $\sigma^+$ PL intensity at a level of 0.4 of the zero field intensity. This is observed for all studied samples (Figures~\ref{fig:fig3}a-d).  

The ability of the dark exciton states to emit both $\sigma^-$ and $\sigma^+$ polarized light also explains the energy shifts of the polarized PL maxima in a magnetic field. The spectral maximum of $\sigma^-$ polarized PL (solid line in Figure~\ref{fig:cartoon}e) is determined by the relation between the $\sigma^-$ polarized emission from the $-2$ (dashed line) and $+2$ (dash-dotted line) states. In a magnetic field, the ZPL emission from the $-2$ state dominates, and the PL maximum shifts toward higher energy. This behaviuor is observed in all studied samples (Figures~\ref{fig:fig3}e-h).

Similarly (Figure~\ref{fig:cartoon}f), the spectral maximum of $\sigma^+$ polarized PL is determined by the emission from the $-2$ and $+2$ states. In high magnetic fields, the 1PL emission from the $-2$ state dominates, and the PL maximum shifts to lower energy.

Overall, the emission from the $-2$ state, $\sigma^-$ from ZPL, and $\sigma^+$ from 1PL, strongly affects the positions of polarized PL maxima. This results in higher energy of $\sigma^-$ polarized PL component (Figures~\ref{fig:fig3}e-h). For comparison, the Zeeman splitting of the dark exciton spin sublevels $-2$ (blue) and $+2$ (red) in a nanocrystal with the c-axis parallel to the magnetic field direction is presented in Figure~\ref{fig:fig3}e by the dashed lines. In addition to the reversed level ordering, the Zeeman splitting of the dark exciton sublevels does not provide the magnitudes of the shifts.

The developed model allows us to describe all unusual features of the polarized PL in CdSe NCs by accounting for the superposition of the ZPL  and 1PL emission of the dark excitons. 

\section{Discussion} From fitting of the magnetic field dependences of the DCP we find that the $g$-factor of the dark exciton $g_F\approx1.7$ is almost independent of the NC diameter. For colloidal CdSe nanocrystals with a diameter of $5.7$~nm, $g_F=1.7$ was determined in Ref.~\cite{Halperin2001}. $g_F=1.3$ was estimated in Ref.~\onlinecite{Granadosdelguila2017} for CdSe NCs with diameters of  $3.5-5$~nm. This value is smaller than $g_F \approx4$ obtained from theoretical estimations \cite{Efros1996} and $g_F=2.7$ from a single CdSe/ZnS nanocrystals study \cite{Biadala2010}. In ESI Section~S4.5$^\dag$ we demonstrate the results of fitting with the assumption of an additional linearly polarized contribution to the ZPL emission, which can be associated with acoustic phonon-assisted recombination of the dark exciton. In this case, we extract $g_F\approx2.5$, almost independent of the NC diameter. Accounting for the 2PL and 3PL emission can further improve the agreement of experiment and theory for $g_F$ and for the energy positions of the $\sigma^+$ polarized PL maximum in the sample D3.3 (Figure~\ref{fig:fig3}h) in low magnetic fields. 

The dark exciton $g$-factor comprises the electron, $g_e$, and hole, $g_h$, $g$-factors as\cite{Efros1996} $g_F=g_e-3g_h$. The size dependence of $g_e$ is well known for CdSe NCs~\cite{Gupta2002,Tadjine2017,Hu2019} (Figure~S25$^\dag$). This allows us to evaluate the hole $g$-factor as $g_h=(g_e-g_F)/3$. Using the $g_F$ values determined from the fitting with the linearly polarized contribution to the ZPL we find $g_h\approx-0.45$ and without it $g_h\approx-0.15$ for NCs with diameters of $4-6$~nm (Figure~S26$^\dag$). It is smaller than the theoretically estimated value of $g_h\approx-1$ from Refs.~\onlinecite{Efros1996,Semina2016}. However, the first value is close to $g_h \approx -0.56$ measured in giant shell CdSe/CdS colloidal nanocrystals from the polarized PL of negative trions.\cite{Liu2013}

The values of the dark exciton $g$-factor used for the modeling of experimental data in Figures \ref{fig:fig2} and \ref{fig:fig3} are obtained in the assumption of thermal equilibrium between the $\pm2$ states. However, the spin relaxation between these states requires a change of the exciton total spin projection by $\Delta F=4$. If the rate of this process is much slower than the dark exciton lifetime, the populations of the $\pm2$ states are determined by the relaxation from the bright $\pm 1$ exciton states. As it is shown in ESI Section~S4.7-8$^\dag$, fitting of the experimental dependences in this case results in larger values of the dark exciton $g$-factor, which are close to the theoretical estimations. The spin relaxation time, or spin dynamics, is often evaluated from the rise of the degree of circular polarization in magnetic field.\cite{Halperin2001,Furis2005} However, such evaluation might be complicated in the case of an ensemble with a nonmonoexponential PL decay.\cite{Shornikova2020NN} We analyze the DCP dynamics in ESI Section~S3$^\dag$, showing that in any case it is faster than the dark exciton recombination and there is not much difference between the time-integrated and saturated DCP values. However, as this time is of the order of the relaxation time from the bright to the dark exciton state (the first time component in the PL decay), it is not possible to conclude, whether the DCP rise and the population of the dark exciton sublevels is determined by the spin relaxation between $\pm1$ or$\pm2$ states. Further clarification of this question is beyond the scope of this paper.

In previous publications, a saturation value of the DCP in high magnetic fields lower than $-0.75$ was associated with nonradiative recombination of the dark excitons,\cite{Halperin2001} partial activation of the dark exciton recombination via admixture of the $0^U$ bright exciton resulting in linearly polarized ZPL emission,\cite{Siebers2015} or the presence of prolate and oblate nanocrystals in the same ensemble.\cite{Liu2014} Accounting for the linearly polarized dark exciton emission assisted by optical phonons is in line with the first two explanations. Here we show, that consideration of the 1PL emission of the dark excitons besides the saturation of the DCP in high magnetic fields determines the DCP spectral dependence. Similar to the samples studied in the present paper, the DCP increase towards the high energy part of the PL spectrum was previously observed for CdSe/CdS dot-in-rods \cite{Siebers2015} and CdTe NCs \cite{Liu2014}. Recently it was also reported for InP NCs \cite{Brodu2019}, where also the strong phonon-assisted emission of dark excitons was observed.  We suggest that the developed theory of the circular polarized PL with account of the phonon-assisted recombination of dark excitons can explain these and other experimental observations for different ensembles of colloidal nanocrystals.

We introduce a phenomenological magnetic field-induced increase of the 1PL recombination rate. This increase is required for the modeling of the experimentally observed $\sigma^-$ and $\sigma^+$ polarized PL maxima positions. Note, that within our model the  $\sigma^-$ and $\sigma^+$ polarized PL maxima from the ensemble have the inverted ordering with respect to the exciton Zeeman splitting, however both shift to higher energies if the 1PL recombination rate is constant in the external magnetic field. Within the second-order perturbation theory, the external magnetic field results only in an increase of the ZPL recombination rate.\cite{Efros1996,Rodina2016} The increase of the 1PL recombination rate in the external magnetic field, as well as the recombination with the assistance of two or three optical phonons, require consideration of higher-order corrections to the recombination rate and will be presented elsewhere. It should be noted that some additional influence on the shift of the polarized PL maxima can come from the F{\"o}rster energy transfer between NCs in dense ensembles.\cite{Furis2005,Liu2014} It results in a spectral diffusion towards lower energy due to the exciton transfer from smaller to larger NCs.  These remarks indicate that the analysis of the polarized PL of dark excitons in ensembles of CdSe nanocrystals can be complicated due to multiple recombination channels with different energy and polarization properties of the emitted light. 

\section{Conclusions}
In summary, the spin polarization of excitons in CdSe NCs embedded in a glass matrix has been studied experimentally in strong magnetic fields up to 30~T. Several unusual features in circularly polarized emission spectra have been found: a low saturation of the degree of circular polarization combined with a pronounced spectral dependence, large and inverted spectral shifts between the oppositely polarized PL components. This puzzling behavior is similar to earlier reports on wet-chemically grown CdSe NCs. We have developed a model that takes into account the cumulative contribution of the zero phonon and one optical phonon-assisted emission of dark excitons to the emission spectra of the NC ensemble. This model describes well all unusual experimental findings and can be readily extended to other colloidal nanocrystals, which inhomogeneous broadening exceeds the optical phonon energy.  

\section{Methods}
\subsection{Samples} The CdSe NCs embedded in glass were synthesized by the following method, which allows considerably reduce the size dispersion of NCs. The batch composition of 61.5 SiO$_{2}$, 15.0 Na$_{2}$O, 10.0 ZnO, 2.5 Al$_{2}$O$_{3}$, 3.0 CdO, 4.1 Se, 2.6 NaF, and 1.3 C (mol.\%) was used to synthesize the initial glass suitable for precipitation of the cadmium selenide crystalline phase. The given amount of activated carbon was introduced directly into the batch in order to provide the required redox conditions of the synthesis. The glass batch was melted in a laboratory electric furnace at $1400-1450^{\circ}$C for 3 hours with stirring for 1 hour at the last stage of melting. The glass melt was poured into graphite molds and annealed in an inertial cooling mode. The initial glass was slightly yellowish in color and optically transparent down to light wavelengths in the near ultraviolet. To isolate the nanostructured CdSe phase, small pieces of the initial glass were heat-treated under isothermal conditions in a two-stage treatment mode, which allows preparation of high quality samples with very low size dispersion of the CdSe nanocrystals (standard deviation $<5$\% as  shown in Ref.~\onlinecite{Golubkov2014}). Precipitation of wurtzite modification of CdSe phase was also evidenced there. Using the data on kinetics of CdSe crystallites growth given in Ref.~\onlinecite{Golubkov2014}, we determined the heat treatment condition to prepare CdSe nanocrystals with different diameters in the range from 3.3~nm up to 6.1~nm. We label the samples with names starting with D followed by their diameter in nanometers. Four samples were studied, which parameters are given in Table~\ref{tab:table1}.

The prepared glass samples with CdSe NCs were examined by small angle X-ray scattering (SAXS) with an infinitely high primary beam using Ni-filtered Cu$K_\alpha$ radiation. We processed the scattering curves, namely the angular dependence of the scattered X-ray intensity, using Guinier plot\cite{Guinier1939} to obtain the radius of gyration, $R_g$, of the CdSe crystallites and evaluate the NC diameters with $D=2.58R_g$.

\subsection{Absorption measurements} Low temperature absorption spectra were measured with an Agilent Cary 6000i UV-Visible-NIR spectrophotometer combined with a helium flow cryostat. The spectra taken at $T=4.2$~K are shown in Figure~S1$^\dag$.

\subsection{Polarized photoluminescence} The samples were placed in the variable temperature insert (2.2~K to 70~K) of a cryostat so that they come into contact with helium exchange gas. External magnetic fields up to 17~T were generated by a superconducting solenoid and applied in the Faraday geometry, i.e. parallel to the direction of photoexcitation and PL collection. The PL was excited with a continuous wave (cw) diode laser (wavelength 405~nm, photon energy 3.06~eV). In all PL experiments the samples were excited nonresonantly, i.e. well above the exciton emission energy. Low excitation densities not exceeding 1~W/cm$^2$ were used to exclude effects caused by exciton-exciton interactions. The PL was detected in backscattering geometry, dispersed with a 0.5 m spectrometer, and measured with a liquid nitrogen-cooled charge-coupled-device camera. 

\subsection{Time-resolved photoluminescence}
A pulsed diode laser (wavelength 405~nm, photon energy 3.06~eV, pulse duration 50~ps, pulse repetition rate 1~MHz) was used for excitation. The PL was dispersed with a 0.5 m spectrometer and its decay was measured with a Si avalanche photodiode connected to a conventional time-correlated single-photon counting module with an overall temporal resolution of 200~ps.

\subsection{Polarization-resolved photoluminescence}
The PL circular polarization degree was analyzed by a combination of a quarter-wave plate and a linear polarizer. Both the magnetic field and spectral dependences of the circular polarization were measured. The degree of circular polarization (DCP) of the PL, $P_c$, is defined by
\begin{equation}
	P_c(t) = \frac { {I^+}{(t)} - {I^-}{(t)}} {{I^+}{(t)}+{I^-}{(t)}}.
\end{equation}
Here, ${I^+}{(t)}$ and ${I^-}{(t)}$ are the intensity of the ${\sigma}^{+}$ and ${\sigma}^{-}$ polarized PL, respectively, measured at time delay $t$ after pulsed excitation. For cw laser excitation, the measured polarization degree corresponds to the time-integrated DCP, ${P_c}^{int}$, which is calculated as
\begin{equation}
	\label{eq:pc}
	P_c^{int} = \frac {\int {I^+}{(t)} dt -\int {I^-}{(t)} dt} {\int {I^+}{(t)} dt + \int {I^-}{(t)} dt}.
\end{equation}

\subsection{Photoluminescence in magnetic fields up to 30~T} The experiments were performed in the High Magnetic Field Lab, Nijmegen. The samples were mounted inside a liquid helium bath cryostat ($T=4.2$~K), which was inserted in a 50~mm bore Florida-Bitter magnet with a maximum dc magnetic field of 32~T. The emission of a 405~nm cw diode laser was focused onto the sample by a lens (10~mm focal length), and the same lens was used to collect the PL. The polarized PL was dispersed with a 0.5 m spectrometer and measured with a liquid nitrogen-cooled charge-coupled-device camera.

\section*{Author contributions}
AAO prepared samples. GQ, EVS and MAP performed the magneto-optical measurements under the guidance of DRY, PCMC and MB. AAG and AVR developed the theory and modeled the data. GQ and EAZ performed the absorption measurements. VKK performed the PLE measurements. IVK and VFS performed the FLN measurements. GQ, AAG and DRY wrote the paper with the assistance of all co-authors. 

\section*{Conflicts of interest}
There are no conflicts to declare.

\section*{Acknowledgements}
We thank Al. L. Efros for valuable discussions and T. V. Shubina for the help with the  PLE measurements. This work was funded by the Deutsche Forschungsgemeinschaft (DFG) in the frame of the International Collaborative Research Centre TRR 160 (Project B1) and the Russian Science Foundation (Grant No. 20-42-01008). IVK and VFS acknowledge partial support by the DFG, TRR 160 (Project B2) and RFBR (Project 19-52-12064) for FLN measurements.  VKK acknowledges partial support by the DFG, TRR 160 (Project C1) and RFBR (Project 19-52-12057) for PLE measurements.  AAG acknowledges support of the Grants Council of the President of the Russian Federation. Experiments in strong magnetic fields up to 30~T were supported by HFML-RU/NWO-I, a member of the European Magnetic Field Laboratory (EMFL).

\bibliographystyle{rsc} %the RSC's .bst file

\clearpage

\pagestyle{empty}

\section*{Supplementary information:  \\Polarized emission of CdSe nanocrystals in magnetic field: the role of phonon-assisted recombination of the dark exciton}

\textit{Gang Qiang, Aleksandr A. Golovatenko, Elena V. Shornikova, Dmitri R. Yakovlev, Anna V. Rodina, Evgeny A. Zhukov, Ina V. Kalitukha, Victor F. Sapega, Vadim Kh. Kaibyshev, Mikhail A. Prosnikov, Peter C. M. Christianen, Aleksei A. Onushchenko, and Manfred Bayer}

%%%%%%%%%% Merge with supplemental materials %%%%%%%%%%
%%%%%%%%%% Prefix a "S" to all equations, figures, tables and reset the counter %%%%%%%%%%
%\pagestyle{plain}
\setcounter{equation}{0}
\setcounter{figure}{0}
\setcounter{table}{0}
\setcounter{page}{1}
%\makeatletter
\renewcommand{\theequation}{S\arabic{equation}}
\renewcommand{\thefigure}{S\arabic{figure}}
\renewcommand{\thetable}{S\arabic{table}}
%\renewcommand{\thepage}{S\arabic{page}}
%\renewcommand{\thesection}{S\arabic{section}}
%%%%%%%%%% Prefix a "S" to all equations, figures, tables and reset the counter %%%%%%%%%%

%\renewcommand{\thefootnote}{\alph{footnote}}

\subsection*{S1. Absorption spectra}\label{SI:absspec} 

\begin{figure*}[hbt]
	\centering
	\includegraphics[width= 17.2 cm]{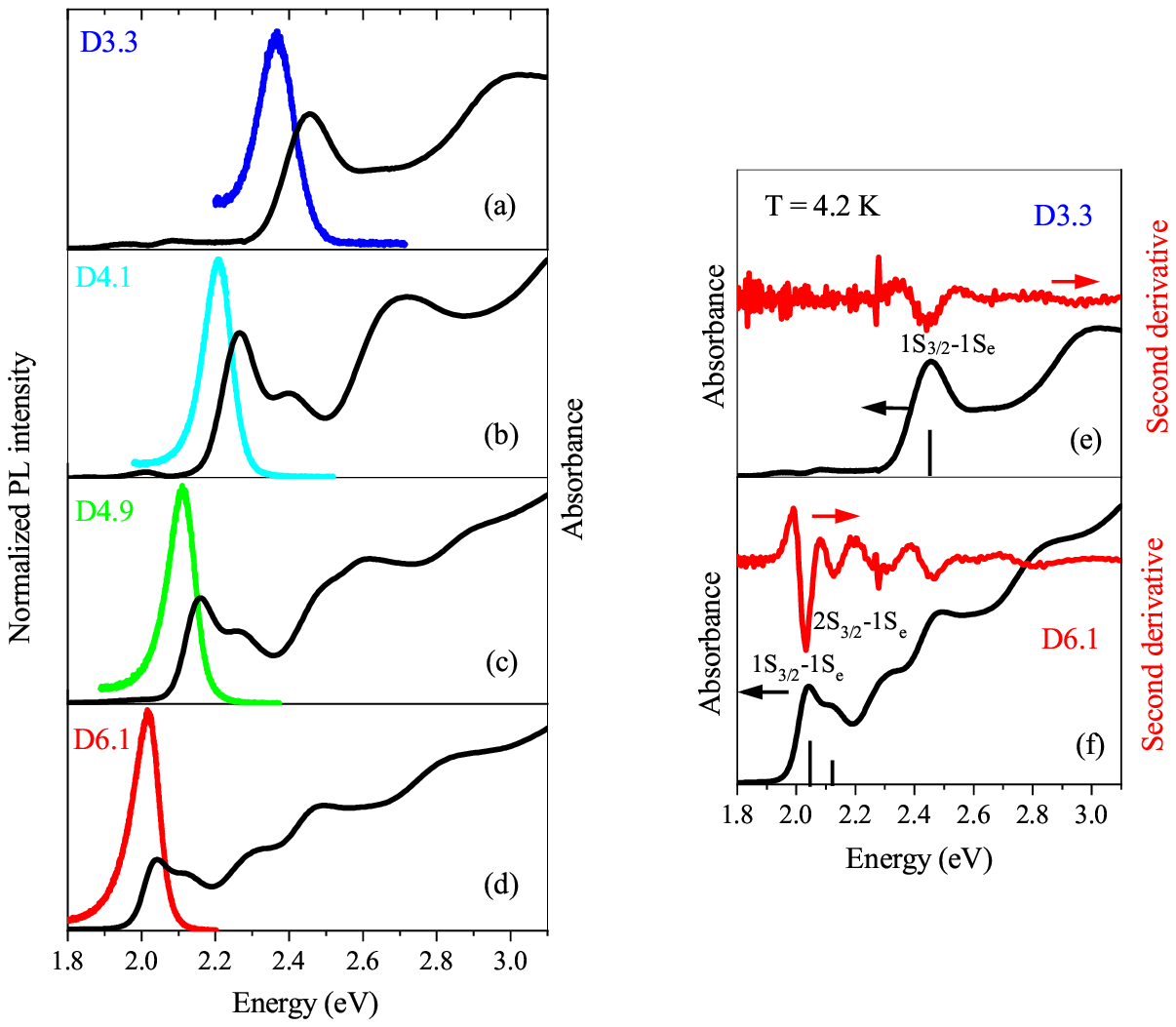} 
	\caption{(a)-(d) PL spectra (color lines) and absorption spectra (black lines) for all studied samples measured at $T=4.2$~K. (e,f) Absorption spectra (black line) and their second derivatives (red line) for D3.3 and D6.1 samples.}  
	\label{fig:figS1}
\end{figure*}

\begin{figure*}[hbt]
	\centering
	\includegraphics[width= 17.2 cm]{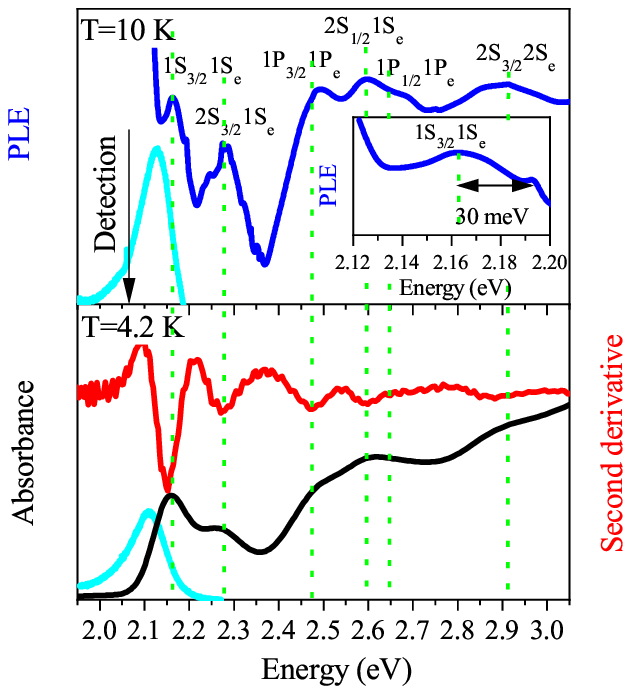} 
	\caption{PLE spectrum at $T=10$~K and absorption spectrum at $T=4.2$~K of sample D4.9. Vertical green dashed lines show the quantum size levels of the exciton. The inset shows a narrow feature shifted by 30 meV from the $1S_{3/2}1S_e$ PLE peak. PL spectra measured under nonresonant excitation are shown by the green solid lines.}
	\label{fig:figSPLE}
\end{figure*}

In Figures~\ref{fig:figS1}(a-d) we show photoluminescence (PL) and absorption spectra for all studied CdSe NCs measured at $T=4.2$~K. One can clearly see that with decreasing the NC diameter from 6.1~nm down to 3.3~nm the PL maximum is shifted to higher energies from 2.017~eV up to 2.366~eV. This shift is provided by the quantum confinement effect. A similar behavior is seen for the lowest peak in the absorption spectra, which corresponds to the lowest bright exciton state. Namely, it shifts from 2.039~eV up to 2.455~eV. The absorption spectra demonstrate a clear modulation corresponding to the optical transitions between different energy levels.\cite{Ekimov1993} This can be better seen when the second derivative of the absorption spectrum is plotted,\cite{Ekimov1993, Park2019} see Figures~\ref{fig:figS1}(e) and \ref{fig:figS1}(f). For example, in sample D6.1 (Figure~\ref{fig:figS1}(f)) the first absorption peak at 2.039~eV is due to the transition between $1S_{3/2}$ and $1S_e$,\cite{NorrisBawendi1996} and the second peak at 2.127~eV can be assigned to the transition between $2S_{3/2}$ and $1S_e$. The length of vertical bars reflects the oscillator strength of the transitions. The parameters of all samples evaluated from the PL and absorption spectra are summarized in Table~\ref{tab:table1}. We also present in Figure~\ref{fig:figSPLE} a comparison between the absorption and photoluminescence excitation (PLE) spectra for the sample D4.9. As one can see, both spectra show maxima near the calculated quantum size levels of the exciton. In the PLE spectrum, several additional narrow features are observed.

\begin{figure*}[hbt]
	\centering
	\includegraphics[width= 17.2 cm]{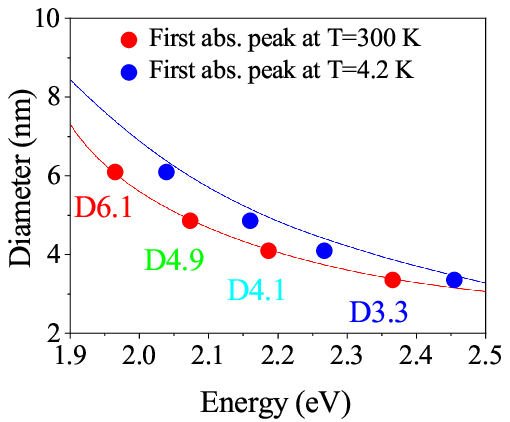} 
	\caption{Exciton absorption energy (the first peak in the absorption spectra) vs NC diameter for the studied CdSe NCs in glass. Symbols are the experimental data points for temperatures of 4.2~K (blue) and 300~K (red). The average NC diameter was determined by SAXS. The red line is an interpolation of the experimental data at $T=300$~K. The blue line corresponds to the experimental data for the $1S_{3/2}1S_e$ absorption peak at $T=10$~K for colloidal bare-core CdSe nanocrystals from Figure 6 in Ref.~\citenum{NorrisBawendi1996}}    
	\label{fig:figdvse}
\end{figure*}

Figure~\ref{fig:figdvse} shows the energy of the first absorption maximum as a function of the NC diameter measured by SAXS (see Methods). The results are given for two temperatures of 4.2 and 300~K. At 4.2~K our results are compared with the experimental data from Figure 6 in Ref.~\onlinecite{NorrisBawendi1996}, obtained at $T=10$~K for CdSe NCs grown by wet chemistry (blue line). 

%\clearpage
\hphantom{\cite{NorrisBawendi1996}}

\subsection*{S2. Bright-dark splitting of excitons in CdSe NCs in glass}\label{SI:dEaf} 

\subsubsection*{S2.1 Theory: temperature dependence of recombination dynamics}\label{SI:tauLtemp}

The exciton fine structure in semiconductor NCs at zero magnetic field is shown schematically in Figure~\ref{fig:fig1}(d). If the system is excited by a short laser pulse, the populations of bright and dark excitons are described by the following rate equations: \cite{Labeau2003}

\begin{equation}\label{eq:nat} 
	\frac{dN_A}{dt} = -{N_A}[\Gamma_A + \gamma_{0}(1 + N_{B})]+ {N_F}\gamma_{0}N_{B},
\end{equation}
\begin{equation} \label{eq:nft} 
	\frac{dN_F}{dt} =-{N_F}(\Gamma_F + \gamma_{0}N_{B})+ {N_A}\gamma_{0}(1 + N_{B}).
\end{equation}
Here, $N_A$ and $N_F$ are the bright and dark exciton populations, respectively.  $\Gamma_{A}$ and $\Gamma_{F}$ are the radiative recombination rates of the bright and dark excitons.  $\gamma_{0}$ is the zero-temperature relaxation rate from the bright $|A\rangle$ to the dark $|F\rangle$ exciton state, it is often referred to as a spin-flip rate. ${{\gamma}_{0}}{N_{B}}$ is the thermal activation rate of the reverse process, where $N_{B} = 1/[\exp(\Delta E_{AF}/k_{B} T)-1]$ is the Bose-Einstein phonon occupation at temperature $T$. $\Delta E_{AF}$ is the bright-dark exciton splitting, $k_B$ is the Boltzmann constant. Under nonresonant excitation, just after the fast initial energy relaxation, the bright and dark exciton states are equally populated: $N_{A}(t=0) = N_{F}(t=0)$. To account for the nonradiative recombination, the PL intensity can be written as:
\begin{equation}\label{eq:It}  
	I(t) = {\eta_A}\Gamma_A{N_A} + {\eta_F}\Gamma_F{N_F} ,
\end{equation}
where $\eta_A$ and $\eta_F$ are the radiative quantum efficiencies of the bright and dark states, respectively. By solving the rate equations (\ref{eq:nat}) - (\ref{eq:It}) in the approximation
$\gamma_0$ $\gg$ $\Gamma_A$ $\gg$ $\Gamma_F$, the radiative recombination dynamics reads as:\cite{Labeau2003}

\begin{equation}\label{eq:Inttime} 
	I(t) = \frac{{\eta_A}\Gamma_A{N_B}+{\eta_F}\Gamma_F}{1+2{N_B}}\exp\left(-\frac{t}{\tau_{Long}}\right)+{\eta_A}\Gamma_A\left[{N_A}(0)-\frac{{N_B}}{1+2{N_B}}\right]\exp\left(-\frac{t}{\tau_{Short}}\right) ,
\end{equation}
\begin{equation}\label{eq:tauLtemp} 
	\tau_{Long}^{-1} = \frac{\Gamma_F + \Gamma_A}{2}-\left(\frac{\Gamma_A - \Gamma_F}{2}\right)\tanh\left(\frac{\Delta E_{AF}}{2{k_B}T}\right) ,
\end{equation}
\begin{equation} \label{eq:taushorttemp} 
	\tau_{Short}^{-1} = \gamma_{0}(1 + 2N_{B}) .
\end{equation}
$\tau_{Long}$ and $\tau_{Short}$ are the time constants corresponding to the long and short component of the PL decay, respectively. The temperature dependence of the PL decay of excitons in neutral NCs can be described by the equations (\ref{eq:Inttime})-(\ref{eq:taushorttemp}). $\Delta E_{AF}$ can be evaluated by fitting an experimentally measured dependence $\tau_{Long}^{-1}(T)$ with equation (\ref{eq:tauLtemp}).

\subsubsection*{S2.2 Experiment: temperature dependence of PL decay and evaluation of $\Delta E_{AF}$}\label{SI:dEaf_exper}

\begin{figure*}[h!]
	\centering
	\includegraphics[width= 17.2 cm]{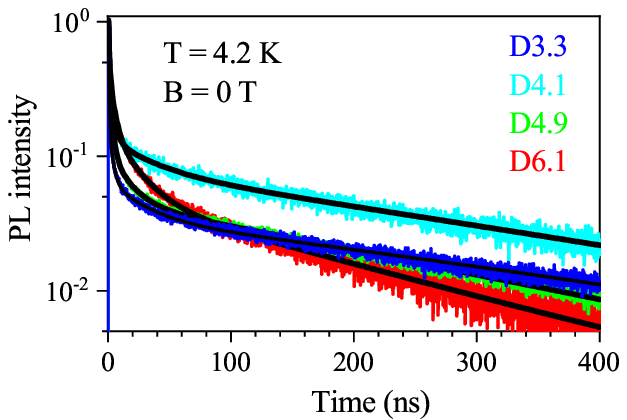} 
	\caption{Photoluminescence decay in CdSe NCs measured at $T=4.2$~K for $B=0$~T. The black lines are fits with four-exponential functions.}  
	\label{fig:figS2}
\end{figure*}

The photoluminescence dynamics for all studied samples measured at $T=4.2$~K for $B=0$~T are shown in Figure~\ref{fig:figS2}. The best fits of the PL decays are achieved with four-exponential functions. The fits are shown by the black lines and the corresponding parameters (times and amplitudes) are listed in Table~\ref{tab:table2}. One can see that the characteristic decay times have about the following values: $T_1$ is on the order of 0.5~ns, $T_2$ on the order of 3~ns, $T_3$ on the order of 25~ns, and $T_4$ on the order of several hundred ns. The shortest time $T_1$ is contributed by the bright exciton recombination and its relaxation to the dark exciton state. The longest time $T_4$ can be assigned to the recombination of dark excitons. 

\begin{table*}[h!]
	\centering
	\small
	\caption{Parameters of the PL decay of the CdSe NCs in glass fitted with four-exponential functions.}
	\begin{tabular*}{0.7\textwidth}{@{\extracolsep{\fill}}lllll}
		\hline
		Sample & D3.3 & D4.1 & D4.9 & D6.1 \\
		\hline
		NC diameter (nm) & 3.3 & 4.1 & 4.9 & 6.1 \\
		Time constant $T_1$ (ns) &  0.4  & 0.5 & 0.5 & 0.8 \\
		Amplitude $A_1$ &  2.1  & 1.4 & 1.8 & 0.9 \\
		Time constant $T_2$ (ns) &  2.8  & 3.3 & 2.3 & 4.6 \\
		Amplitude $A_2$ &  0.2  & 0.2 & 0.4 & 0.3 \\
		Time constant $T_3$ (ns) &  27  & 29 & 15 & 23 \\
		Amplitude $A_3$ &  0.03  & 0.06 & 0.07 & 0.11 \\
		Time constant $T_4$ (ns) &  410  & 315 & 245 & 170 \\
		Amplitude $A_4$ &  0.04  & 0.08 & 0.05 & 0.06 \\
		\hline
	\end{tabular*}
	\label{tab:table2}
\end{table*}

\begin{figure*}[h!]
	\centering
	\includegraphics[width= 17.2 cm]{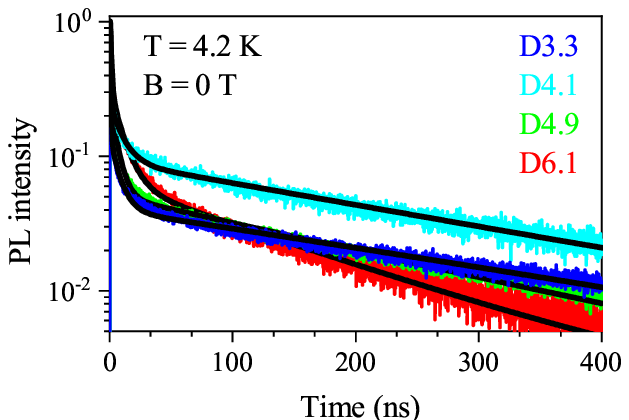} 
	\caption{Photoluminescence decay in CdSe NCs measured at $T=4.2$~K for $B=0$~T. The black lines are fits with three-exponential functions.}  
	\label{fig:figS2s}
\end{figure*}

The PL decay can be also reasonably well fitted with three-exponential functions. We show these results for comparison in Figure~\ref{fig:figS2s} and give the parameters in Table~\ref{tab:table3}. One can see that the $T_1$ times are similar in both cases. The $T_2$ time in the three-exponential fit lies in between the $T_2$ and $T_3$ times in the four-exponential fit. The longest $T_3$ time in the three-exponential fit is about $20-25$\% shorter than the $T_4$ time in the four-exponential fit. For the evaluation of $\Delta E_{AF}$ from the temperature dependence of the longest time, we use the data from the four-exponential fit.

\begin{table*}[h!]
	\centering
	\small
	\caption{Parameters of the PL decay of CdSe NCs in glass fitted with three-exponential functions.}
	\begin{tabular*}{0.7\textwidth}{@{\extracolsep{\fill}}lllll}
		\hline
		Sample & D3.3 & D4.1 & D4.9 & D6.1 \\
		\hline
		NC diameter (nm) & 3.3 & 4.1 & 4.9 & 6.1 \\
		Time constant $T_1$ (ns) &  0.5  & 0.7 & 0.6 & 1.0 \\
		Amplitude $A_1$ &  2.0  & 1.3 & 1.6 & 1.0 \\
		Time constant $T_2$ (ns) &  7.4  & 9.8 & 5.7 & 9.8 \\
		Amplitude $A_2$ &  0.1  & 0.2 & 0.2 & 0.3 \\
		Time constant $T_3$  (ns) &  316  & 263 & 208 & 137 \\
		Amplitude $A_3$ &  0.04  & 0.09 & 0.05 & 0.06 \\
		\hline
	\end{tabular*}
	\label{tab:table3}
\end{table*}

\begin{figure*}[hbt]
	\centering
	\includegraphics[width= 17.2 cm]{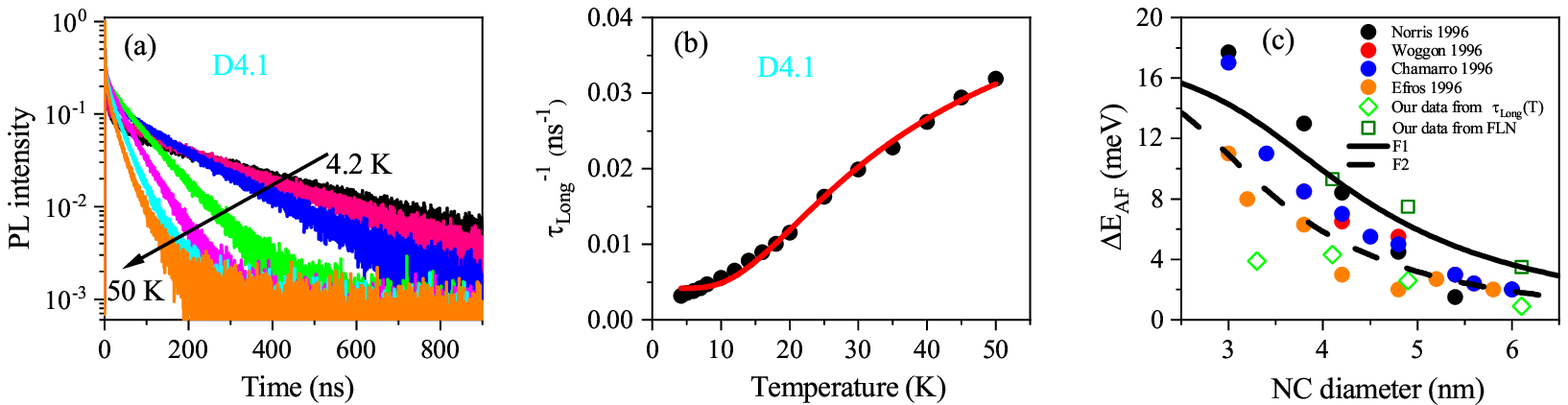} 
	\caption{(a) PL decay in sample D4.1 at various temperatures increasing from 4.2~K up to 50~K. (b) Temperature dependence of the decay rate for the long component $\tau^{-1}_{Long}$ in D4.1. The red line shows a fit with equation~(\ref{eq:tauLtemp}) for $\Delta E_{AF}=4.3$~meV. (c) Size dependence of the bright-dark splitting $\Delta E_{AF}$. The green diamonds are the data from $\tau_{Long}(T)$ dependence. The green open squares are the data determined from FLN. Literature data for wet chemistry grown and glass-embedded CdSe NCs are shown by the solid circles.\cite{Norris1996, Woggon1996, Chamarro1996, Efros1996} Size dependences of $\Delta E_{AF}$ calculated with  $\eta=0.1(a_B/a)^3$~meV and  $\eta=0.2(a_B/a)^3$~meV for spherical wurtzite CdSe nanocrystals are shown by dashed and solid lines, respectively. }   
	\label{fig:figS3}
\end{figure*}

Figure~\ref{fig:figS3}(a) shows a representative example of the PL dynamics variations with increasing temperature from 4.2 up to 50~K. It is given for the sample D4.1 at $B=0$~T. The long tail of the PL decay shortens with increasing temperature, i.e. the recombination rate $\tau_{Long}^{-1}$ accelerates. The temperature dependence of $\tau_{Long}^{-1}$ is plotted in Figure~\ref{fig:figS3}(b) by dots. The fit by equation~(\ref{eq:tauLtemp}) is shown by the red line. From the fit  we evaluate the bright-dark splitting $\Delta E_{AF}=4.3$~meV. The $\Delta E_{AF}$ values for the other studied samples are given in Table~\ref{tab:table1} and shown by the open diamonds in Figure~\ref{fig:figS3}(c). They are compared with the literature data available for the wet chemistry grown and glass-embedded  CdSe NCs, see the closed circles in Figure~\ref{fig:figS3}(c). Good agreement is found for $\Delta E_{AF}$ in D6.1, D4.9 and D4.1, while for D3.3, $\Delta E_{AF}$ deviates from the general tendency.

Lines in Figure~\ref{fig:figS3}(c) show size dependences of $\Delta E_{AF}$ calculated as: $\Delta E_{\rm AF}=2\eta+\Delta/2-(4\eta^2+\Delta^2/4-\eta\Delta)^{1/2}$,\cite{Efros1996,Goupalov} where $\eta\propto(a_B/a)^3$ is a measure of the electron-hole exchange interaction, $\Delta=23$~meV is the crystal field spitting in spherical wurtzite CdSe NCs of the hole states with angular momentum projections $\pm3/2$ and $\pm1/2$, $a_B=5.6$~nm is the exciton Bohr radius in bulk CdSe, $a$ is the NC radius. The dashed line is calculated with accounting for the short-range exchange interaction (but not the long-range) between electron and hole\cite{Efros1996} corresponding to $\eta=0.1(a_B/a)^3$~meV. The solid line fits $\Delta E_{AF}$ determined from FLN (see next section) and it is calculated with $\eta=0.2(a_B/a)^3$~meV. Note, that the full account of both, short-range exchange and long-range exchange interaction \cite{Goupalov} would correspond to $\eta=0.37(a_B/a)^3$~meV.

%\clearpage
\subsubsection*{S2.3 Evaluation of $\Delta E_{AF}$ from fluorescence line narrowing}\label{SI:FLN}

An alternative method of $\Delta E_{AF}$ determination is fluorescence line narrowing, which also allows us to compare the intensities of the ZPL and 1PL emission lines of the dark exciton. For resonant excitation of the CdSe NCs we used a ring dye laser with R6G dye and a dye laser with DCM dye. The scattered light was analyzed by a Jobin-Yvon U1000 double monochromator equipped with a cooled GaAs photomultiplier. To record sufficiently strong signals of the dark exciton emission and to suppress the laser stray light, a spectral slit width of 0.2 cm$^{-1}$ (0.025 meV) was used. The measurements were performed on samples immersed in pumped liquid helium (typically at a temperature of 2 K). The FLN signal was measured in backscattering geometry with linearly polarized excitation (H) and subsequent detection of the PL with orthogonal linear polarization (V). We evaluate $\Delta E_{AF}=3.5$, 7.5 and 8.4 meV  in the samples D6.1, D4.9 and D4.1, respectively. The relative intensity of the ZPL emission is larger in nanocrystals with a smaller diameter.   

\begin{figure*}[hbt]
	\centering
	\includegraphics[width= 17.2 cm]{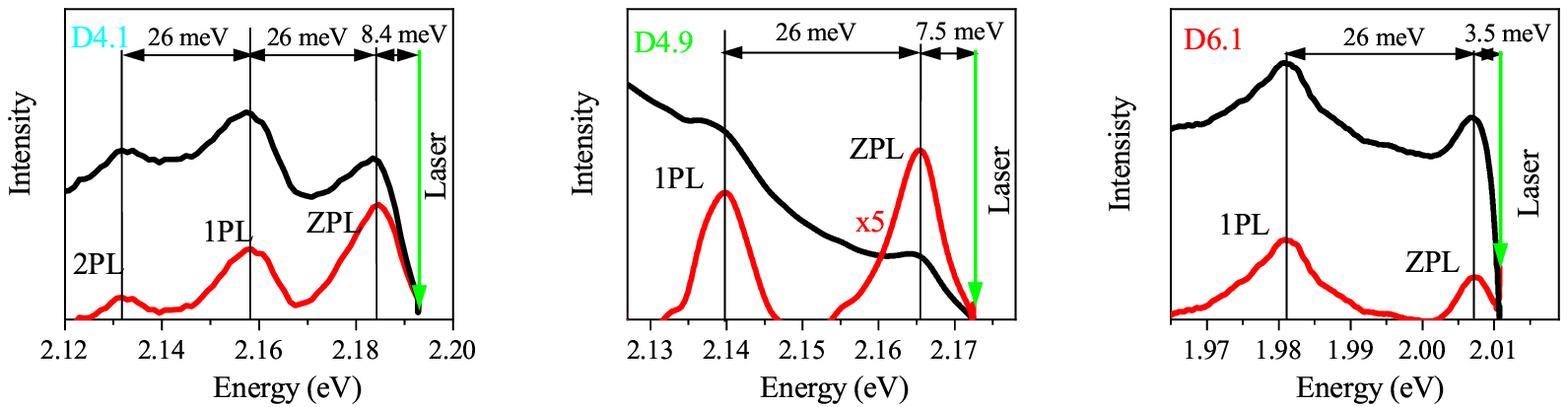} 
	\caption{Fluorescence line narrowing spectra of samples D4.1, D4.9 and D6.1. The black lines show the FLN spectra as recorded. The red lines show the ZPL and 1PL emission after subtraction of the nonresonant PL background. The black arrows show the energies of the optical phonon $E_{LO}=26\pm0.5$~meV and the bright-dark exciton splittings $\Delta E_{AF}(D4.1)=8.4$~meV, $\Delta E_{AF}(D4.9)=7.5$~meV and $\Delta E_{AF}(D6.1)=3.5$~meV. }
	\label{fig:fln}
\end{figure*}

\subsection*{S3. Degree of circular polarization (DCP) dynamics}\label{SI:spinrelax} 

The magnetic field induced circular polarization is determined by the exciton population of the Zeeman spin levels split in a magnetic field. The equilibrium DCP, $P_c^{eq}$, is determined only by the thermal distribution, i.e. by the ratio of the exciton Zeeman splitting to the thermal energy $k_B T$, while the time-integrated DCP, $P_c^{int}$,  is also affected by the spin-relaxation and recombination processes.\cite{Liu2014} It is because when the spin relaxation time is longer than the exciton lifetime, the thermal equilibrium is not established, which results in $P_c^{int} < P_c^{eq}$.  The relationship between them can be described by $P_c^{int}(B)=d P_c^{eq}(B)$, where $0<d\le1$ is the so-called dynamical factor. In case of a single exponential decay $d=\tau/(\tau+\tau_s)$, where $\tau$ is the exciton lifetime and $\tau_s$ is the exciton spin relaxation time. For a multi-exponential decay, an averaging should made, accounting for the times and relative amplitudes of each component, e.g. see Ref.~\onlinecite{Shornikova2020NN}. For the present study it is important to ensure that the $P_c^{int}$ measured under cw excitation is close to $P_c^{eq}$, i.e., that $d \approx 1$. 

\begin{figure*}[hbt]
	\centering
	\includegraphics[width= 17.2 cm]{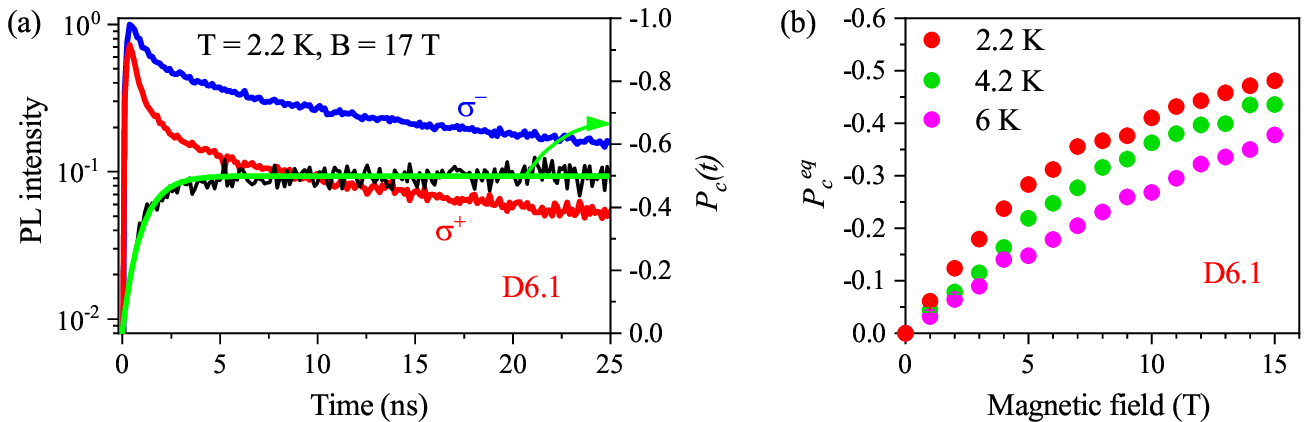}
	\caption{(a) Polarized PL decay for sample D6.1 measured for $T=2.2$~K and $B=17$~T. Red and blue lines correspond to $\sigma^{+}$ and $\sigma^{-}$ polarizations, respectively. Black line shows  the time-resolved DCP. (b) Magnetic field dependence of equilibrium DCP ($P_c^{eq}$) at $T=2.2$, 4.2 and 6~K for the sample D6.1.}  
	\label{fig:figS4}	
\end{figure*}

The polarization-resolved PL decay is shown in  Figure~\ref{fig:figS4}(a) for the sample D6.1. The $\sigma ^{-}$ and $\sigma ^{+}$ polarized components are measured at $T=2.2$~K and $B=17$~T, and the corresponding time-resolved DCP $P_c (t)$ is shown by the black line. After pulsed laser excitation, the DCP changes rapidly within a nanosecond from 0 to about $-0.35$, and then slowly evolves until saturating at $-0.49$, which corresponds to the equilibrium DCP, $P_c ^{eq}$. $P_c (t)$ can be described by the following empirical expression: \cite{Liu2013}
\begin{equation}\label{eq:pct}
	P_c(B, t) =   P_c^{eq}(B)[1- \exp(-t/ {\tau_{s}(B)})] ,
\end{equation}
where $\tau_s$ characterize the DCP rise time but is not  the true spin relaxation time in general case. $P_c ^{eq}$ is strongly influenced by the magnetic field and temperature, see Figure~\ref{fig:figS4}(b). It increases with increasing magnetic field and is suppressed by elevated temperature. At $T=2.2$~K, $P_c ^{eq}$ varies from 0 up to $-0.48$ at $B=15$~T, when the temperature increases, it drops to $-0.44$ at $T=4.2$~K and to $-0.38$ at $T=6$~K.

\begin{figure*}[hbt]
	\centering
	\includegraphics[width= 17.2 cm]{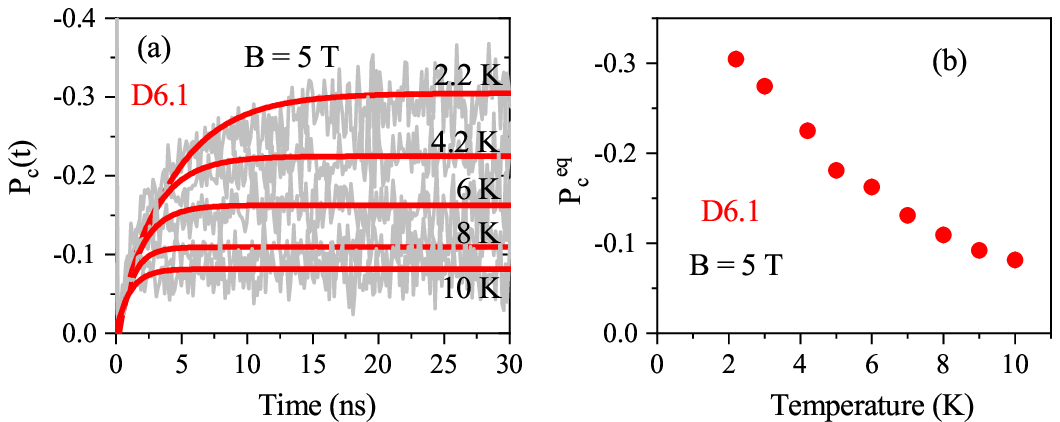}
	\caption{(a) Temperature dependence of the time-resolved DCP, $P_c (t)$, at $B=5$~T, shown by grey curves. Red lines are fits with equation~(\ref{eq:pct}). (b) Temperature dependence of the equilibrium DCP, $P_c^{eq}$, at $B=5$~T for sample D6.1.}
	\label{fig:figS5}	
\end{figure*}

Figure~\ref{fig:figS5}(a) shows the time-resolved DCP measured at $B=5$~T for various temperatures, which can be well reproduced by equation~(\ref{eq:pct}), see the solid red lines. With increasing  temperature, the DCP rise accelerates, while the equilibrium DCP, $P_c^{eq}$, decreases ($-0.30$ at $T=2.2$~K and $-0.08$ at $T=10$~K) as shown in Figure~ \ref{fig:figS5}(b).

%\clearpage

\begin{figure*}[hbt]
	\centering
	\includegraphics[width= 17.2 cm]{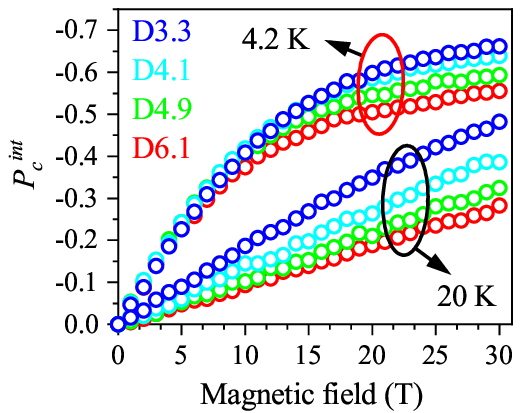}
	\caption{Magnetic field dependence of the $P_c^{int}$ at $T=4.2$~K and 20~K for all samples.}
	\label{fig:fig4S_pcint4K20K}	
\end{figure*}

Figure~\ref{fig:fig4S_pcint4K20K} shows the magnetic field dependence of $P_c^{int}$ at $T=20$~K and 4.2~K for comparison. The $P_c^{int}$ at $T=20$~K is smaller than the one at liquid helium temperature and changes almost linearly with magnetic field.

\begin{figure*}[h!]
	\centering
	\includegraphics[width= 17.2 cm]{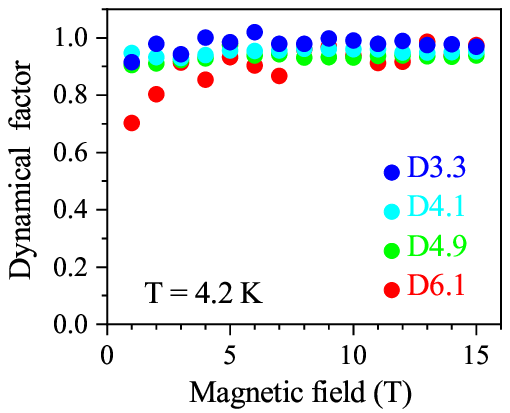}
	\caption{Magnetic field dependence of the dynamical factor $d$ measured at $T=4.2$~K.}
	\label{fig:figS6}	
\end{figure*}

The magnetic field dependence of the dynamical factor $d(B)=P_c^{int}(B)/P_c^{eq}(B)$ in all studied samples is shown in Figure~\ref{fig:figS6}.At $T=4.2$~K it is close to unity. Therefore, the time-integrated DCP is close to the value expected for the equilibrium exciton polarization. 

%\clearpage

\begin{figure*}[hbt]
	\centering
	\includegraphics[width= 17.2 cm]{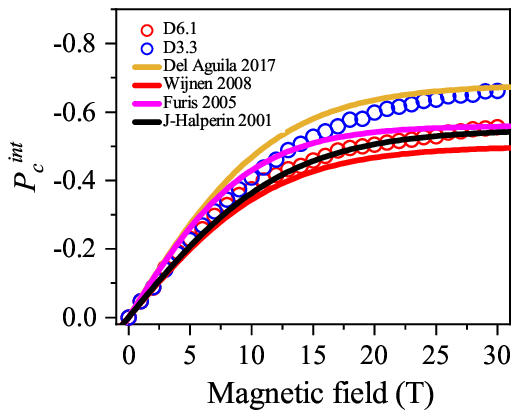}
	\caption{Magnetic field dependences of the DCP at $T=4.2$~K in CdSe nanocrystals from the present study (open circles) and literature data: red line from Ref.~\citenum{Wijnen2008} for 3.6~nm CdSe/CdS(1ML) NCs, black line from Ref.~\citenum{Halperin2001} for 5.7~nm CdSe NCs, magenta line from Ref.~\citenum{Furis2005} for 2.6~nm CdSe/ZnS NCs (average between DCP at $T=1.6$~K and 10~K), yellow line from Ref.~\citenum{Granadosdelguila2017} for CdSe and CdSe/CdS NCs with diameters of $3.5-5$~nm.}
	\label{fig:dcpALL}
\end{figure*}

In Figure~\ref{fig:dcpALL}, we present a comparison of the magnetic field dependences of the time-integrated DCP measured in this work with the dependences from Refs.~\onlinecite{Halperin2001,Furis2005,Wijnen2008,Granadosdelguila2017}. One can see that in all cases a similar increase of the DCP in low magnetic fields is observed, while the saturation of the DCP at $B=30$~T varies significantly. For glass embedded CdSe NCs the saturation value of the DCP in high magnetic fields increases with decreasing NC diameter. For the colloidal CdSe NCs studied in Refs.~\onlinecite{Halperin2001,Furis2005,Wijnen2008,Granadosdelguila2017} there is no obvious dependence of the DCP saturation on the NC diameter.         

\hphantom{\cite{Wijnen2008,Halperin2001,Furis2005,Granadosdelguila2017}}

%\clearpage
\subsection*{S4. Model description} 
\label{SI:theory}

\subsubsection*{S4.1 ZPL and 1PL emission}\label{SI:zpl1pl}

\begin{figure*}[h!]
	\centering
	\includegraphics[width= 17.2 cm]{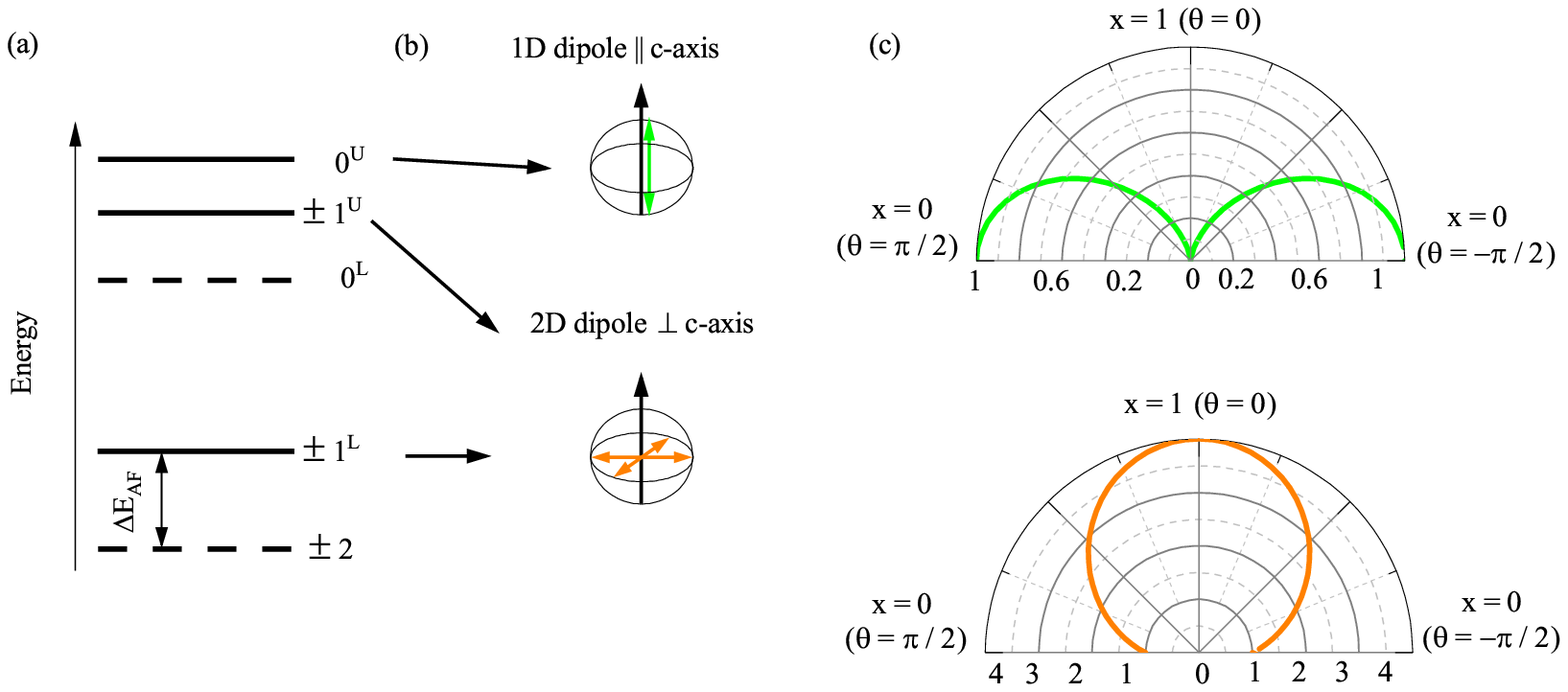}
	\caption{(a) Fine structure of the $1S_{3/2}1S_e$ exciton. Dashed lines show the dark excitons $\pm2$ and $0^L$. Solid lines show the bright excitons $\pm1^L$, $\pm1^U$ and $0^U$. (b) The emission of the $0^U$ bright exciton has the properties of a 1D dipole, while the emission of the $\pm1^L$ and $\pm1^U$ bright excitons has the properties of a 2D dipole. (c) Spatial distribution of the emission intensity for the $\pm1^L$, $\pm1^U$ excitons (orange line) and the $0^U$ exciton (green line) with respect to the direction of the c-axis.\cite{Efros1996}
	}
	\label{fig:figFS}	
\end{figure*}

To describe the experimental dependences of the DCP and the positions of the $\sigma^+$ and $\sigma^-$ polarized PL maxima, we extend the theoretical model of the circularly polarized emission from an ensemble of randomly oriented nanocrystals, first developed in Ref.~ \onlinecite{Halperin2001}. The extension accounts for the linearly polarized contribution coming from the dark exciton recombination assisted by optical phonons. The  model in Ref.~ \onlinecite{Halperin2001} accounted for only the zero phonon line (ZPL) emission of the $\pm2$ dark excitons via the admixture of the $\pm1^L$ bright excitons. This admixture results in the recombination of the $\pm2$ dark excitons with the emission of photons that are circularly polarized in the plane perpendicular to the c-axis of a nanocrystal (Figure~\ref{fig:figFS}).\cite{Efros1996,Rodina2016} The spatial distribution of the PL intensity, in this case, corresponds to the emission of a 2D dipole and has a maximum for the propagation of light along the direction of the c-axis (Figure~\ref{fig:figFS}(c)). The model in Ref.~\onlinecite{Halperin2001} does not consider the contribution of the optical phonon-assisted recombination of the dark excitons.

However, it is well known from fluorescence line narrowing experiments, that the dark exciton PL also has a strong contribution from optical phonon-assisted recombination.\cite{Norris1996, Woggon1996} In the model described below, we consider the recombination of dark excitons with the assistance of one optical phonon, the so-called 1PL emission. The 1PL emission is arranged predominantly via admixture of the $0^U$ bright exciton.\cite{Rodina2016} This results in the recombination of $\pm2$ dark excitons with the emission of photons that are linearly polarized along the c-axis of a nanocrystal.\cite{Efros1996,Rodina2016}  The spatial distribution of the PL intensity, in this case, corresponds to the emission of the 1D dipole and has a maximum for the propagation of light in the plane perpendicular to the c-axis (Figure~\ref{fig:figFS}(c), green curve). Also, we analyze the impact of the dark exciton recombination via the admixture of the $0^U$ bright exciton, caused by the interaction with acoustic phonons.\cite{Rodina2016} The inhomogeneous broadening of the PL spectra in the samples under investigation is much larger than the typical energies of the acoustic phonons. For this reason, we neglect the energy shift between the ZPL emission and the acoustic phonon-assisted emission, i.e. we consider the emission at the ZPL energy which includes the contributions coming from the admixture of the $0^U$ and $\pm1^L$ bright excitons to the dark exciton.       

With the account of the ZPL and 1PL contributions and the random orientation of the NCs in an ensemble, the spectral dependences of the $\sigma^{\pm}$ polarized PL intensities in a magnetic field are given by:
\begin{equation} \label{IEB}
	I^\pm(E,B) = \int_0^1 dx \sum_{i=\pm 2} \left[ I_{i,ZPL}^\pm(x,B) f_{ZPL}(E-\delta E_i(B,x))+ I_{i,1PL}^\pm(x,B) f_{1PL}(E-\delta E_i(B,x)) \right] \, ,	
\end{equation}
where $\delta E_{\pm 2}(B,x)=\pm g_F \mu_BB x/2$ are the Zeeman shifts of the $\pm 2$ dark excitons, $\mu_B$ is the Bohr magneton, $g_F$ is the dark exciton $g$-factor, $x=\cos \theta$ with $\theta$ being the angle between the c-axis of the nanocrystal and the magnetic field direction. $I_{i,ZPL}^\pm(x,B)$  and $ I_{i,1PL}^\pm(x,B)$ are the intensities of the dark exciton emission with $\sigma^+$ ($\sigma^-$) polarization in the external magnetic field applied in the Faraday geometry for NCs without and with emission of an optical phonon, respectively. 

The functions $f_{ZPL}(E)$ and $f_{1PL}(E)$ determine the inhomogeneous broadening of the PL spectra due to the  NC size dispersion in an ensemble. We consider the case of a normal distribution:  
\begin{align}
	&f_{ZPL}(E-\delta E_i(B,x)) = \frac{1}{w\sqrt{2\pi}} \exp\left( -\frac{(E-\delta E_i(B,x)-E_{ZPL}^0)^2}{2w^2}\right)	 \, , \\
	& f_{1PL}(E-\delta E_i(B,x)) = \frac{1}{w\sqrt{2\pi}} \exp\left( -\frac{(E-\delta E_i(B,x)-E_{1PL}^0)^2}{2w^2}\right)	 =f_{ZPL}(E-\delta E_i(B,x)+E_{LO}) \nonumber .
\end{align}
Here $E_{LO}=26$~meV is the optical phonon energy in CdSe, $E_{ZPL}^0$ and $E_{1PL}^0=E_{ZPL}^0-E_{LO}$ are the maxima of the ZPL and 1PL emission from the dark excitons, corresponding to the maximum of the size distribution.

The experimental data on the DCP of the emission presented in this paper are measured under continuous-wave excitation. Thus, we are interested in the time-integrated PL intensities $I_{i,ZPL}^\pm(x,B)$ and $I_{i,1PL}^\pm(x,B)$, which are defined as follows:
\begin{align}
	&	I^{\pm}_{\pm 2,ZPL}(x,B)=\Gamma^{\pm}_{\pm 2,ZPL}(x,B)N_{ex}(x,B), \quad I^{\pm}_{\pm 2,1PL}(x,B)=\Gamma^{\pm}_{\pm 2,1PL}(x,B)N_{ex}(x,B), \\
	&	N_{ex}(x,B)=\frac{G}{\gamma_{tot}(x,B)} ,
\end{align}
where $G$ is the generation rate of excitons, $\gamma_{tot}$ is the total exciton recombination rate, $N_{ex}$ is the equilibrium population of the excitons.  

The external magnetic field not only splits the exciton Zeeman sublevels (by the field component parallel to the $c$-axis), but also mixes the bright to the dark exciton states (by the field component perpendicular to the $c$-axis). This mixing results in the additional activation of the dark exciton, which can be described within second order perturbation theory in moderate magnetic fields \cite{Efros1996,Rodina2016}. The resulting  magnetic field and angular dependent recombination  rates $\Gamma^{\pm}_{\pm 2,ZPL}(x,B)$  contributing to the ZPL emission are given by:
\begin{align}
	&\Gamma_{-2,ZPL}^{\pm}(x,B)=\frac{1}{2}\gamma_{\varepsilon}n_{-2}(x,B)\left[\left(\left(\frac{g_e\mu_BB}{2\sqrt{2}\varepsilon}\right)^2(1-x^2)+1\right)(1\mp x)^2+\chi_{ac}(1-x^2)\right], \\
	&\Gamma_{+2,ZPL}^{\pm}(x,B)=\frac{1}{2}\gamma_{\varepsilon}n_{+2}(x,B)\left[\left(\left(\frac{g_e\mu_BB}{2\sqrt{2}\varepsilon}\right)^2(1-x^2)+1\right)(1\pm x)^2+\chi_{ac}(1-x^2)\right], \\
	&\gamma_{\varepsilon}=\frac{\varepsilon^2}{6\eta^2}\frac{1}{\tau_0},\quad \eta=0.2\left(\frac{a_{B}}{a}\right)^3~{\rm meV} .
\end{align}
where $g_e$ is the electron $g$-factor, $\varepsilon$ is the characteristic energy of interaction, which results in the admixture to the dark states of the $\pm1^L$ bright exciton states in zero magnetic field, $\chi_{ac}=\gamma_{ac}/\gamma_{\varepsilon}$ is the ratio of the linearly and circularly polarized recombination rates of the ZPL emission at $B=0$~T, $\tau_0$ is the lifetime of the $0^U$ bright exciton, $n_{\pm2}(x,B)$ are the Boltzmann populations of the $\pm2$ dark exciton states, respectively. Note that the angular dependence ($x$-dependence) arises not only from the perpendicular component of the magnetic field, but is also caused by the spatial profiles of the emission distribution for the 1D and 2D dipoles. In modeling we use $\eta=0.2(a_B/a)^3$~meV which fits $\Delta E_{AF}$ values from FLN measurements. The only parameter in our modeling which depends on $\eta$ is $r_{lin}$ determining polarization properties of 1PL emission (see below). For $\eta=0.1(a_B/a)^3$~meV  corresponding to short-range exchange interaction between electron and hole, relative change of $r_{lin}$ does not exceed 2\% for the whole range of considered NC diameters,  and does not affect our conclusions about the role of the 1PL emission of the dark excitons.  

The angle-specific recombination rates of the dark excitons with assistance of the optical phonons are given by:
\begin{align}
	&\Gamma_{-2,1PL}^{\pm}(x,B)=\frac{1}{2}\gamma_{\varepsilon}\chi(x,B) n_{-2}(x,B)\left[r_{lin}(1-x^2)+(1-r_{lin})(1\mp x)^2\right], \\
	&\Gamma_{+2,1PL}^{\pm}(x,B)=\frac{1}{2}\gamma_{\varepsilon}\chi(x,B) n_{+2}(x,B)\left[r_{lin}(1-x^2)+(1-r_{lin})(1\pm x)^2\right],\\
	&\chi(x,B)=\chi_0\left[1+c_{1PL}\left(\frac{g_e\mu_BB}{2\sqrt{2}\varepsilon}\right)^2(1-x^2)\right] .
\end{align}
Here $\chi_0=\gamma_{\rm LO}/\gamma_{\varepsilon}$ is the ratio of the 1PL recombination rate to the ZPL recombination rate at $B=0$~T. $c_{1PL}$ is a phenomenological parameter, which determines the increase of the 1PL recombination rate in a magnetic field. The parameter $r_{lin}$ determines the fraction of the optical phonon-assisted recombination rate via the admixture to the $0^U$ bright exciton. According to Ref.~\onlinecite{Rodina2016} it equals to:  
\begin{align}
	r_{lin}=\left(\frac{E_{LO}^2(\Delta+4\eta+E_{LO})^2}{2(3\Delta\eta+E_{LO}(\Delta+4\eta)+E_{LO}^2)^2}+1\right)^{-1}. 
\end{align}
The factor 2 in the denominator instead of 4 in equation (19) from Ref.~\onlinecite{Rodina2016} is used because here we are interested in recombination rates, while in Ref.~\onlinecite{Rodina2016} transition dipole moments were compared.

The total recombination rate equals to:
\begin{align}
	\gamma_{tot}(x,B)=\gamma_{\varepsilon}\left[\left(\frac{g_e\mu_BB}{2\sqrt{2}\varepsilon}\right)^2(1-x)^2+1+\chi(x,B)+\chi_{ac}\right].
\end{align}

\subsubsection*{S4.2 Ensemble-averaged PL decay}\label{SI:decayfit}

The fitting of the DCP and the positions of the polarized PL maxima is performed together with fitting of the $\tau_{Long}(B)$ dependence. As the PL decay is measured from an ensemble of randomly oriented nanocrystals, we need to develop a procedure for averaging the exciton lifetime over the ensemble. Let us consider the decay of the normalized PL intensity of the dark excitons at a fixed energy $E$ for the randomly oriented ensemble. The increase of the dark exciton recombination rate depends on the angle $\theta$ between the c-axis of the NC and the direction of the magnetic field. The total PL decay equals to the sum of decays from NCs having all possible values of the angle $\theta$:
\begin{align}
	& I(E,B,t)=A_{\rm L} \frac{\int_0^1dxG(x,B)\exp\left[-t\tilde \gamma_{tot}(x,B)\right]}{
		\int_0^1dxG(x,0)}\, , \\ 
	&\tilde \gamma_{tot}(x,B)=\frac{\gamma_{tot}(x,B)}{\gamma_{tot}(x,0)\tau_{B=0}} =\frac{\gamma_{tot}(x,B)}{(\gamma_{\varepsilon}+\gamma_{LO}+\gamma_{ac})\tau_{B=0}}\, , \\
	&G(x,B)=f_{ZPL}(E)\sum_{i=\pm2}\Gamma_{i,ZPL}^{\pm}(x,B)+f_{1PL}(E)\sum_{i=\pm2}\Gamma_{ i,1PL}^{\pm}(x,B).
\end{align} 
Here $I(E,B,0)$ is PL intensity at $t=0$, $A_{\rm L}$ is the amplitude of the decay component corresponding to the dark exciton recombination in the total PL decay. Note that $\gamma_{tot}(x,0)=\gamma_{\varepsilon}+\gamma_{LO}+\gamma_{ac}$ is the radiative recombination rate of the dark excitons in zero magnetic field and $\tau_{B=0}$ is the lifetime of the dark excitons in zero magnetic field. We neglected the contribution of the Zeeman splitting of the dark excitons in the functions $f_{ZPL}(E)$ and $f_{1PL}(E)$, as these energies are much smaller than the characteristic PL linewidth. The characteristic lifetime determining the PL decay of the dark excitons in magnetic field for a randomly oriented ensemble of NCs can be calculated as:
\begin{equation} \label{tau_ens}
	\tau_{\rm ens}(B)= \frac{\int_0^1 dx \,  G(x,B) \tilde\gamma_{tot}(x,B)^{-1}    }{ \int_0^1 dx \, G(x,B)  } \, .
\end{equation}
It has the meaning of an ensemble-averaged dark exciton lifetime. The comparison of the calculated lifetimes $\tau_{\rm ens}(B)$ (curves) and the experimental times $\tau_{Long}(B)$ (symbols) is presented in Figure~\ref{fig:fig1}(f).  The fitting parameters are $\varepsilon$, $\chi_0$, $\chi_{ac}$, and $c_{1PL}$, see Tables~\ref{tab:tablefit} and \ref{tab:tablefitAC}.  These parameters are fixed by fitting the $\tau_{Long}(B)$ dependences and used in the further fitting of the field dependences of the PL intensities and PL maxima.

\subsubsection*{S4.3 Energy of PL maximum in zero magnetic field}\label{SI:PLmax}

In an inhomogeneous ensemble of NCs the energy of the PL maximum at $B=0$~T is determined by the relative intensities of the ZPL and 1PL emission of the dark excitons. One can find the energy of the PL maximum using the following consideration. At $B=0$~T the PL intensity is given by:     
\begin{equation} \label{IE}
	I^\pm(E) = \int_0^1 dx \sum_{i=\pm 2} \left[ I_{i,ZPL}^\pm(x) f_{ZPL}(E)+ I_{i,1PL}^\pm(x) f_{1PL}(E) \right] \, .	
\end{equation}
In zero magnetic field the averaging over the random angle distribution of the ensemble in Eq.~\eqref{IE} results in the same intensity for both circular polarizations $I^+(E)=I^-(E)$, as expected.

We can rewrite Eq.~\eqref{IE} in zero magnetic field as 
\begin{equation} \label{IE0}
	I^\pm(E,0) = f_{ZPL}(E) \left< I_{ZPL} \right>+ f_{1PL}(E)  \left< I_{1PL} \right> \, ,	
\end{equation}
where
\begin{align}
	\left< I_{ZPL} \right> = \int_0^1 dx \sum_{i=\pm 2} I_{i,ZPL}^\pm(x,0) =\frac{4+2\chi_{ac}}{3(1+\chi_0)} , \\ \left< I_{1PL} \right> = \int_0^1 dx \sum_{i=\pm 2}  I_{i,1PL}^\pm(x,0)=\frac{2\chi_0(2-r_{lin})}{3(1+\chi_0)} \, .
\end{align}
Taking the derivative of Eq.~\eqref{IE0} we find the following condition for the PL maximum $E_{\rm max}$ as
\begin{equation}\label{Emax}
	E_{\rm max} = E_{ZPL}^0-\frac{E_{LO}}{1+\beta_0 f(E_{\rm max}) }=E_{ZPL}^0-E_{LO}{\tilde \beta_0(E_{\rm max} )}\, , 
\end{equation}
where $\beta_0=\left< I_{ZPL} \right>/\left< I_{1PL} \right>=(2+\chi_{ac})/\chi_0(2-r_{lin})$,  $\tilde \beta_0=(1+\beta_0f(E_{\rm max}))^{-1}$ and  
\begin{equation}\label{fEmax}
	f(E_{\rm max}) = \frac{f_{ZPL}(E_{\rm max} )}{ f_{1PL}(E_{\rm max} ) }  =\exp\left[\frac{E_{LO}^2(1-2\tilde \beta_0(E_{\rm max}))}{2w^2}\right]\, . 
\end{equation}
It should be noted that the parameter $\beta_0$ allows us to find the relation between $\chi_{ac}$ and $\chi_{0}$ in zero magnetic field. The parameter $\beta_0$ can be determined from the comparison of the ZPL and 1PL intensities measured in fluorescence line narrowing experiments.    

\begin{figure*}[h!]
	\centering
	\includegraphics[width= 17.2 cm]{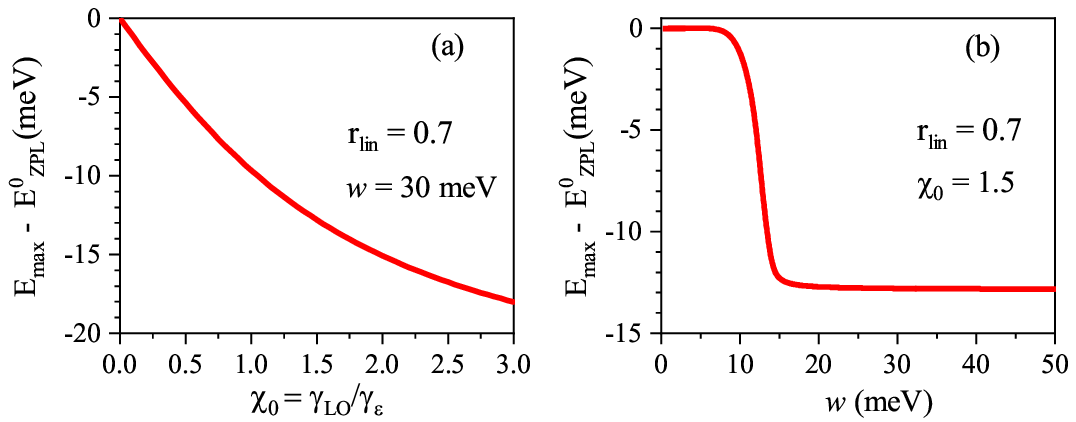}
	\caption{Energy of the PL maximum $E_{\rm max}$ as a function of (a) the LO-assisted recombination rate and (b) the inhomogeneous broadening of PL spectrum.}
	\label{fig:figS8}	
\end{figure*}

In Figure~\ref{fig:figS8}(a) the dependence of $E_{\rm max}$ on the  parameter $\chi_0$ is presented with $\chi_{ac}=0$, $r_{\rm lin}=0.7$ and $w=30$~meV. One can see that an increase of $\chi_0$, representing the relative intensity of the 1PL emission, results in a shift of $E_{\rm max}$ towards the maximum of the 1PL emission. For a fixed $\chi_0=1.5$ one can see in Figure~\ref{fig:figS8}(b) that a sufficiently large inhomogeneous broadening ($w>15$~meV) results in a constant shift of $E_{\rm max}$ from $E_{\rm ZPL}^0$. For $w<15$~meV a sharp return of $E_{\rm max}$ to $E_{\rm ZPL}^0$ is observed. The reason is that the spectra of the ZPL and 1PL emission do not overlap anymore. In this case one can separate the ZPL and 1PL peaks, as usually observed in fluorescence line narrowing experiments.  

\subsubsection*{S4.4 Calculation of PL maximum energy in magnetic field }\label{SI:PLmaxMF}

To calculate the energies of the $\sigma^+$ and $\sigma^-$ polarized PL maxima, presented in Figures~\ref{fig:fig3}(e-h), we use the following approach. 
Taking the derivative of Eq.~\eqref{IEB}, we find a condition for the total maxima of the $\sigma^\pm$ polarized PL in magnetic field, $E_{\rm max}^\pm(B)$.  We assume that the Zeeman energies $\delta E_i$ and the maximum shifts in magnetic field  $\Delta E_{\rm max}^\pm(B) = E_{\rm max}^\pm(B) - E_{\rm max}$ are much smaller than $w$ and keep only the linear-in-magnetic field corrections to the energy terms and to the distribution functions, to write the following conditions for $\Delta E_{\rm max}^\pm(B)$:
\begin{eqnarray}\label{EmaxB}
	\Delta E_{\rm max}^\pm(B) &=& \frac{g_F\mu_B B }{2} \frac{\left< \delta I^\pm_{\rm ZPL} (B )\right> f_0 (1-E_{LO}^2{\tilde \beta_0}^2/w^2) +
		\left< \delta I^\pm_{1PL} (B )\right>  [1-E_{\rm LO}^2(1-{\tilde \beta_0})^2/w^2]  }{\left<  I^\pm_{\rm ZPL} (B )\right> f_0 (1-E_{LO}^2{\tilde \beta_0}^2/w^2) +
		\left<  I^\pm_{1PL} (B )\right>  [1-E_{LO}^2(1-{\tilde \beta_0})^2/w^2] } \nonumber \\
	&+& E_{LO} \frac{ {\tilde \beta}_0  (f_0 \beta_\pm (B) + 1)-1 }{f_0 \beta_\pm (B) (1-E_{LO}^2{\tilde \beta_0}^2/w^2)  +  [1-E_{LO}^2(1-{\tilde \beta_0})^2/w^2]   } \, .
\end{eqnarray}
Here,  
\begin{equation}
	\left< I_{ZPL}^\pm(B) \right> = \int_0^1 dx \sum_{i=\pm 2} I_{i,ZPL}^\pm(x,B)  \, ,	\quad \left< I_{1PL} ^\pm(B) \right> = \int_0^1 dx \sum_{i=\pm 2}  I_{i,1PL}^\pm(x,B) \, ,
\end{equation}
\begin{eqnarray}
	\left<\delta I_{ZPL}^\pm(B) \right> = \int_0^1 dx [ I_{+2,ZPL}^\pm(x,B) -  I_{-2,ZPL}^\pm(x,B) ]  \, ,	\\  
	\left< \delta  I_{1PL} ^\pm(B) \right> = \int_0^1 dx  [I_{+2,{1PL}}^\pm(x,B) -  I_{-2,{1PL}}^\pm(x,B)]  \, ,
\end{eqnarray}
and $\beta_\pm(B) = \left< I_{ZPL}^\pm(B) \right>/\left< I_{1PL}^\pm(B) \right>$. The second term in Eq.~\eqref{EmaxB} depends on the magnetic field only via $\beta_\pm(B)$.  Note that $\beta_\pm (0)=\beta_0$ and both terms in Eq.~\eqref{EmaxB} vanish at $B=0$.

%\clearpage
\subsubsection*{S4.5 Fitting parameters}
\label{SI:fitresults}

The calculated results presented in Figures~\ref{fig:fig1}f, \ref{fig:fig2} and \ref{fig:fig3} were achieved assuming that the ZPL emission is governed solely by the admixture of the $\pm1^L$ bright exciton. The fit parameters for this scenario are given in Table~\ref{tab:tablefit}.    

\begin{table*}[h!]
	\centering
	\small
	\caption {\ Fitting parameters in Figures~\ref{fig:fig1}f, \ref{fig:fig2} and \ref{fig:fig3}.}
	\begin{tabular*}{0.8\textwidth}{@{\extracolsep{\fill}}llllll}
		\hline
		Sample & D3.3 & D4.1 & D4.9 & D6.1& \\
		\hline
		$g_F$ & 1.6 & 1.8 & 1.8 & 1.6&   best fit\\
		$g_e$ & 1.42 & 1.32 & 1.23 & 1.1& Refs.~\onlinecite{Gupta2002,Tadjine2017,Hu2019} \\
		$g_h$ & $-0.06$ & $-0.16$ & $-0.19$ & $-0.16$& $g_h=(g_e-g_F)/3$ \\
		$\varepsilon$ (meV) &  0.23  &0.24 & 0.25 &0.32&  best fit \\
		$\chi_0=\gamma_{LO}/\gamma_{\varepsilon}$ &  0.8  & 1 & 1.3 & 1.5&  best fit \\
		$\chi_{ac}=\gamma_{ac}/\gamma_{\varepsilon}$&0&0&0&0&\\
		$c_{1PL}$ &  0.2  & 0.35 & 0.45 & 0.5 &  best fit\\
		$w$ (meV) & 50 & 36 & 32& 30 & PL linewidth\\
		\hline
	\end{tabular*}
	\label{tab:tablefit}
\end{table*}

\begin{figure*}[h!]
	\centering
	\includegraphics[width= 17.2 cm]{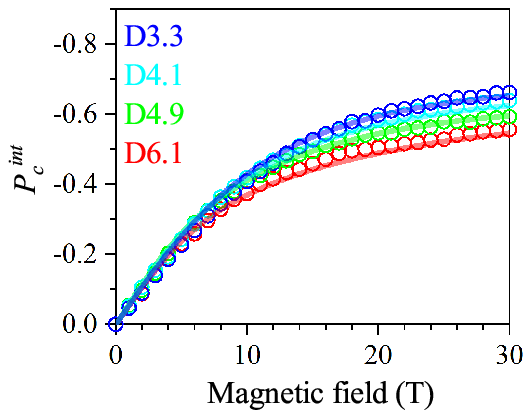}
	\caption{Magnetic field dependences of $P_c^{int}(B)$ measured at the PL maximum in all samples under study. Lines are fits with Eq.~(\ref{eq:pcint}). The contribution to the ZPL emission via admixture of the $0^U$ bright exciton is taken into account.}	
	\label{fig:fig3ac}
\end{figure*}

\begin{figure*}[h!]
	\centering
	\includegraphics[width= 17.2 cm]{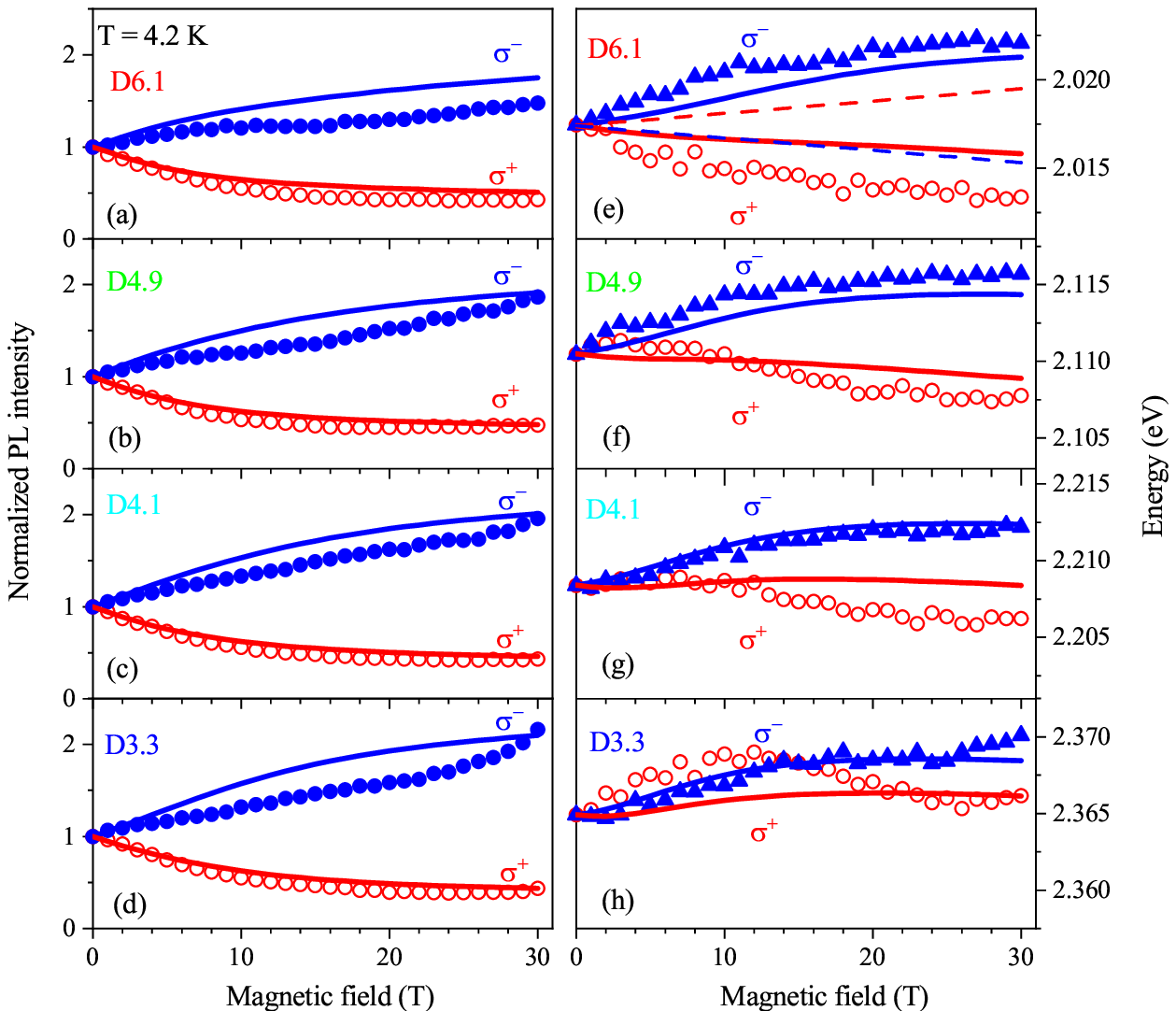}
	\caption{PL intensity and spectral shifts in magnetic field. (a-d) Time-integrated intensity of the ${\sigma}^{+}$ (red) and ${\sigma}^{-}$ (blue) polarized PL as function of the magnetic field in CdSe NCs. (e-h) Corresponding PL peak energies. For all panels, the symbols correspond to the experimental data, while curves show the results of calculations. In the calculations, the contribution to the ZPL emission via the admixture of the $0^U$ bright exciton is taken into account. Dashed lines in panel (e) show the Zeeman splitting of the dark exciton spin sublevels $-2$ (blue) and $+2$ (red) in a nanocrystal with c-axis parallel to the magnetic field direction and $g_F=2.4$.}	
	\label{fig:figshiftAC}
\end{figure*}

We also considered the case, in which the ZPL emission contains a contribution from the recombination of the dark excitons via the admixture of the $0^U$ bright exciton. For all studied samples we assumed that the recombination rates of the dark excitons through the admixture of the $\pm1^L$ and $0^U$ bright excitons are equal, i.e. $\chi_{ac}=\gamma_{ac}/\gamma_{\varepsilon}=1$. The fit results in this case are presented in Figures~\ref{fig:fig3ac} and \ref{fig:figshiftAC}. Within this scenario we also observe a good agreement with the experimental data. The main result that we obtain from the inclusion of the dark exciton recombination via the admixture of the $0^U$ bright exciton is an increase of the $g$-factor $g_F\approx2.5$ which is close to the value reported for single CdSe NCs.\cite{Biadala2010} The dark exciton $g$-factors and other fitting parameters for this scenario are given in Table~\ref{tab:tablefitAC}. 

\begin{table*}[h!]
	\centering
	\small
	\caption {\ Fitting parameters in Figures \ref{fig:fig3ac},  \ref{fig:figshiftAC} and \ref{fig:spectrDCPAC}.}
	\begin{tabular*}{0.8\textwidth}{@{\extracolsep{\fill}}llllll}
		\hline
		Sample & D3.3 & D4.1 & D4.9 & D6.1& \\
		\hline
		$g_F$ &2.4 & 2.6 & 2.6 & 2.4&best fit \\
		$g_e$ & 1.42 & 1.32 & 1.23 & 1.1& Refs.~\onlinecite{Gupta2002,Tadjine2017,Hu2019} \\
		$g_h$ & $-0.32$ & $-0.42$ & $-0.45$ & $-0.43$& $g_h=(g_e-g_F)/3$\\
		$\varepsilon$ (meV) &  0.18  &0.19 & 0.19 &0.25&best fit \\
		$\chi_0=\gamma_{LO}/\gamma_{\varepsilon}$ & 1  & 1.4 & 1.75 & 2.25&best fit \\
		$\chi_{ac}=\gamma_{ac}/\gamma_{\varepsilon}$&1&1&1&1&\\
		$c_{1PL}$ &  0.12  & 0.18 & 0.25 & 0.25& best fit \\
		$w$ (meV) & 50 & 36 & 32& 30& PL linewidth \\
		\hline
	\end{tabular*}
	\label{tab:tablefitAC}
\end{table*}

\clearpage
\subsubsection*{S4.6 Spectral dependence of DCP at $B=30$~T}\label{SI:specDCP}

\begin{figure*}[h!]
	\centering
	\includegraphics[width= 17.2 cm]{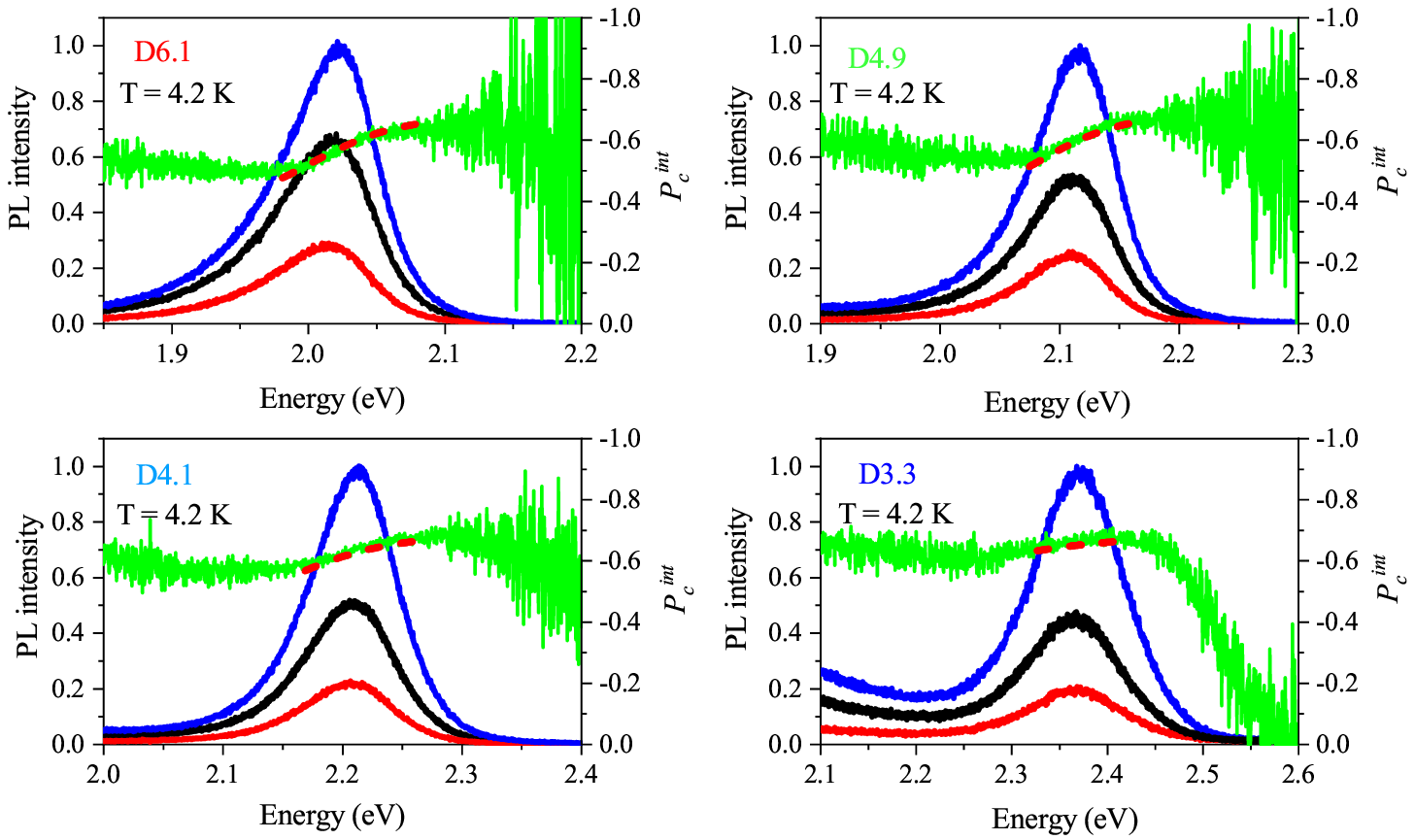}
	\caption{PL spectra of the ${\sigma}^{+}$ (red) and ${\sigma}^{-}$ (blue) polarized components at $B=30$~T and PL spectrum at $B=0$~T (black). Green line shows the experimental spectral dependence of the circular polarization degree at $B=30$~T. Red dashed line shows the calculated spectral dependence of the DCP at $B=30$~T with accounting for the ZPL emission via the admixture of the $\pm1^L$ bright excitons solely. }
	\label{fig:spectrDCP}
\end{figure*}

\begin{figure*}[h!]
	\centering
	\includegraphics[width= 17.2 cm]{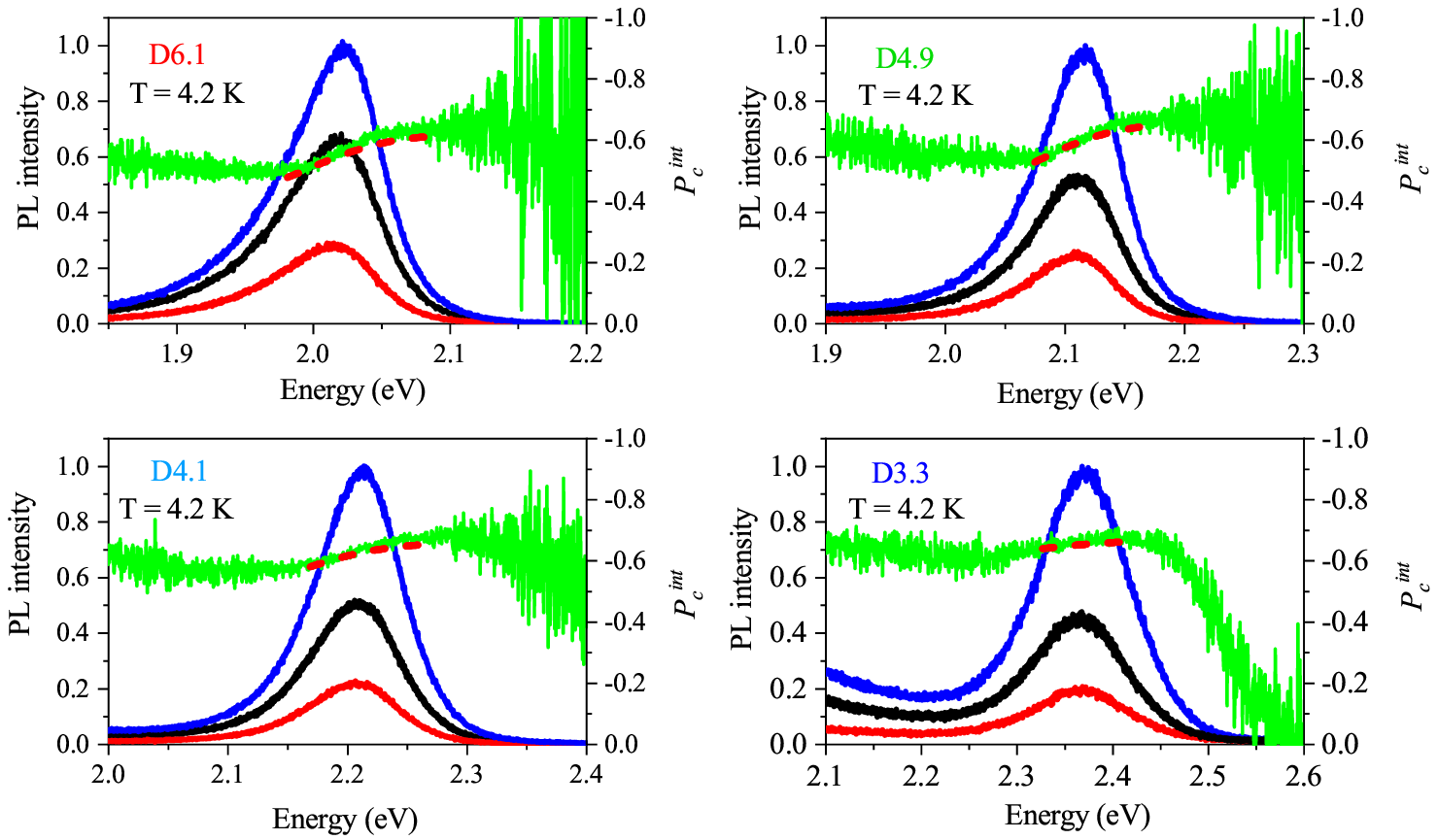}
	\caption{PL spectra of the ${\sigma}^{+}$ (red) and ${\sigma}^{-}$ (blue) polarized components at $B=30$~T and PL spectrum at $B=0$~T (black). Green line shows the experimental spectral dependence of the circular polarization degree at $B=30$~T. Red dash line shows the calculated spectral dependence of the DCP at $B=30$~T. The emission of the dark excitons at the ZPL energy via the admixture of the $0^U$ bright exciton is included.}
	\label{fig:spectrDCPAC}
\end{figure*}

Here we present a comparison of the calculated and experimental spectral dependences of the DCP for all studied CdSe NCs at $B=30$~T. The results of the calculation for the ZPL emission governed only by the admixture of the bright exciton $\pm1^L$ with the fitting parameters from Table \ref{tab:tablefit} are presented in Figure~\ref{fig:spectrDCP}. The results of the calculation with additinal inclusion of the ZPL emission governed by the admixture of the bright exciton $0^U$ with the fitting parameters from Table \ref{tab:tablefitAC} are presented in Figure~\ref{fig:spectrDCPAC}. From the fit results we conclude that the increase of the degree of the PL polarization towards higher energies is governed by the decrease of the 1PL contribution. This effect is pronounced in large NCs, while in the sample D3.3 the DCP is almost constant across the PL spectrum.

\clearpage
\subsubsection*{S4.7 Fitting for non-equilibrium population of $\pm2$ states}

\begin{figure*}[h!]
	\centering
	\includegraphics[width= 17.2 cm]{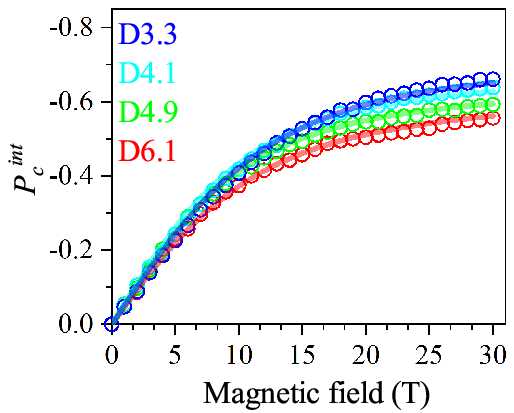}
	\caption{Magnetic field dependences of the $P_c^{int}(B)$ measured at the PL maximum for all samples. Lines are fits with Eq.~(\ref{eq:pcint}). Only the contribution to the ZPL emission via the admixture of the $\pm1^L$ bright exciton is taken into account. The populations of the dark exciton states $\pm2$ are considered to be determined by the relaxation from the $\pm1^L$ states. }	
	\label{fig:figdcpga}
\end{figure*}

\begin{figure*}[h!]
	\centering
	\includegraphics[width= 17.2 cm]{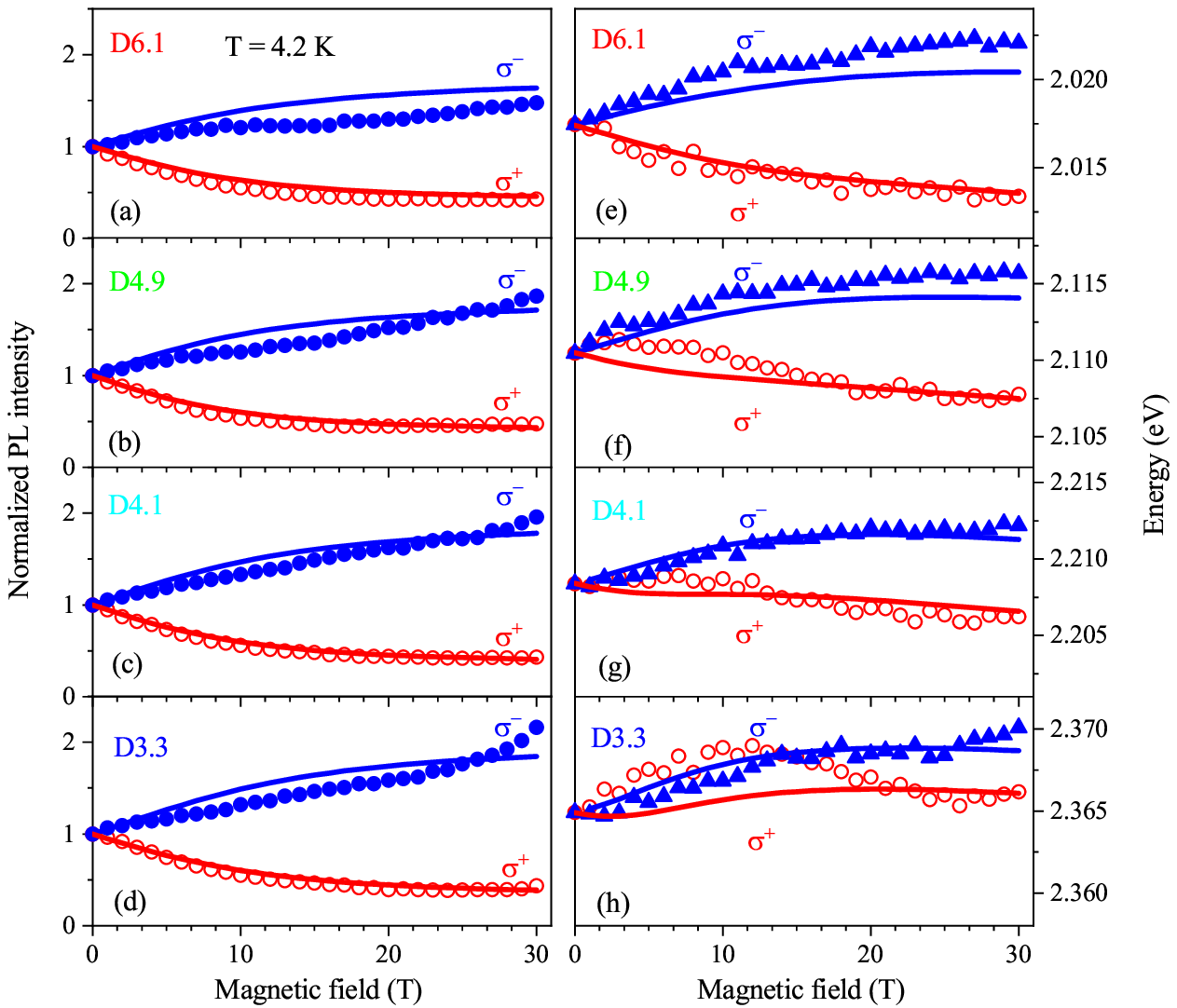}
	\caption{PL intensity and spectral shifts in magnetic field. (a-d) Time-integrated intensity of the ${\sigma}^{+}$ (red) and ${\sigma}^{-}$ (blue) polarized PL as function of the magnetic field in CdSe NCs. (e-h) Magnetic field dependences of the corresponding PL peak energies. For all panels the symbols correspond to the experimental data, while the curves show the results of calculations. Only the contribution to the ZPL emission via the admixture of the $\pm1^L$ bright exciton is taken into account. The populations of the dark exciton states $\pm2$ are considered to be determined by the relaxation from the $\pm1^L$ states.}	
	\label{fig:figshiftga}
\end{figure*}

\begin{figure*}[h!]
	\centering
	\includegraphics[width= 17.2 cm]{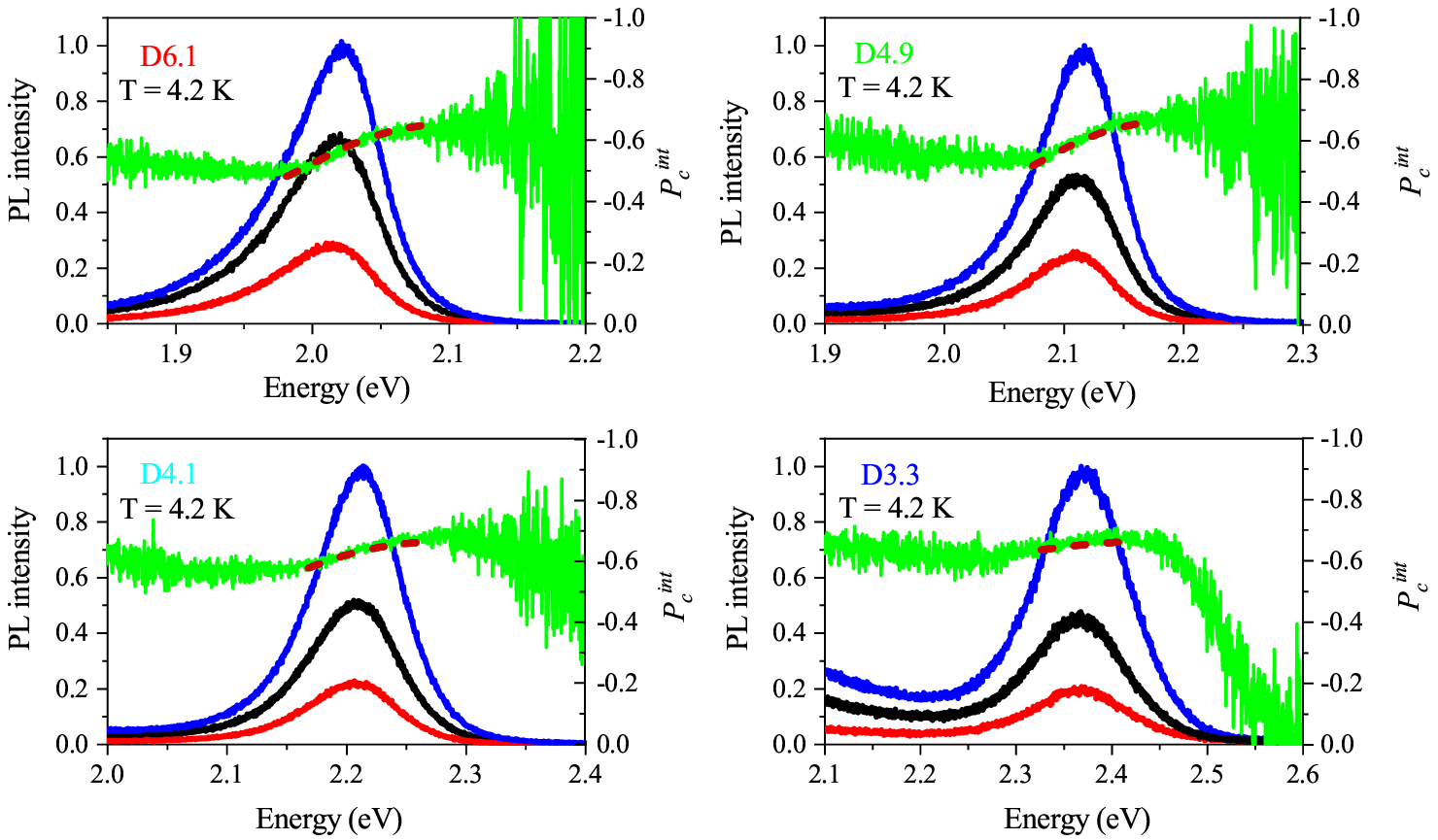}
	\caption{ PL spectra of the ${\sigma}^{+}$ (red) and ${\sigma}^{-}$ (blue) polarized components at $B=30$~T and PL spectrum at $B=0$~T (black). Green line shows the experimental spectral dependence of the circular polarization degree at $B=30$~T. Red dashed line shows the calculated spectral dependence of the DCP at $B=30$~T with accounting for the ZPL emission via the admixture of the $\pm1^L$ bright excitons solely. The populations of the dark exciton states $\pm2$ are considered to be determined by the relaxation from the $\pm1^L$ states. }
	\label{fig:spectrDCPga}
\end{figure*}

Here we consider the case when the populations of the $\pm2$ states due to slow spin relaxation do not approach the equilibrium values in the applied magnetic field. We assume that after nonresonant excitation the excitons relax to the $\pm1^L$ states which are split in the applied magnetic field. We also assume that before the relaxation to the $\pm2$ states thermal equilibrium between the $\pm1^L$ states is achieved. Then excitons from the $-1^L$ state relax to the $-2$ state, while excitons from the $+1^L$ state relax to the $+2$ state. In this case, the populations of excitons in the $\pm2$ states are determined by the $g$-factor of the bright exciton, while the splitting of the $\pm2$ states is determined by the $g$-factor of the dark exciton. The relationship  between the bright exciton $g$-factor and the $g$-factors of electron and hole is given in Ref.~\onlinecite{Efros2003}. From fitting of the experimental data (Figures \ref{fig:figdcpga}, \ref{fig:figshiftga}, \ref{fig:spectrDCPga})  without the acoustic phonon-assisted contribution to the ZPL, we find $g_h=-0.5$ for D6.1; $-0.6$ for D4.9; $-0.77$ for D4.1; and $-0.43$ for D3.3, with all the other fitting parameters being the same as in Table~\ref{tab:tablefit}. 

%\clearpage

\begin{figure*}[h!]
	\centering
	\includegraphics[width= 17.2 cm]{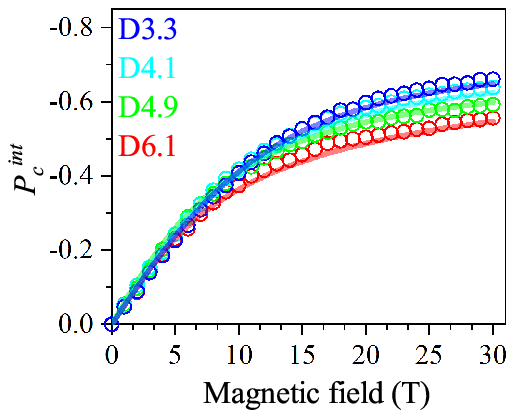}
	\caption{Magnetic field dependences of the $P_c^{int}(B)$ measured at the PL maximum in all samples. Lines are fits with Eq.~(\ref{eq:pcint}). The contribution to the ZPL emission via the admixture of the $0^U$ bright exciton is taken into account. The populations of the dark exciton states $\pm2$ are considered to be determined by the relaxation from the $\pm1^L$ states. }	
	\label{fig:figdcpgaac}
\end{figure*}

\begin{figure*}[h!]
	\centering
	\includegraphics[width= 17.2 cm]{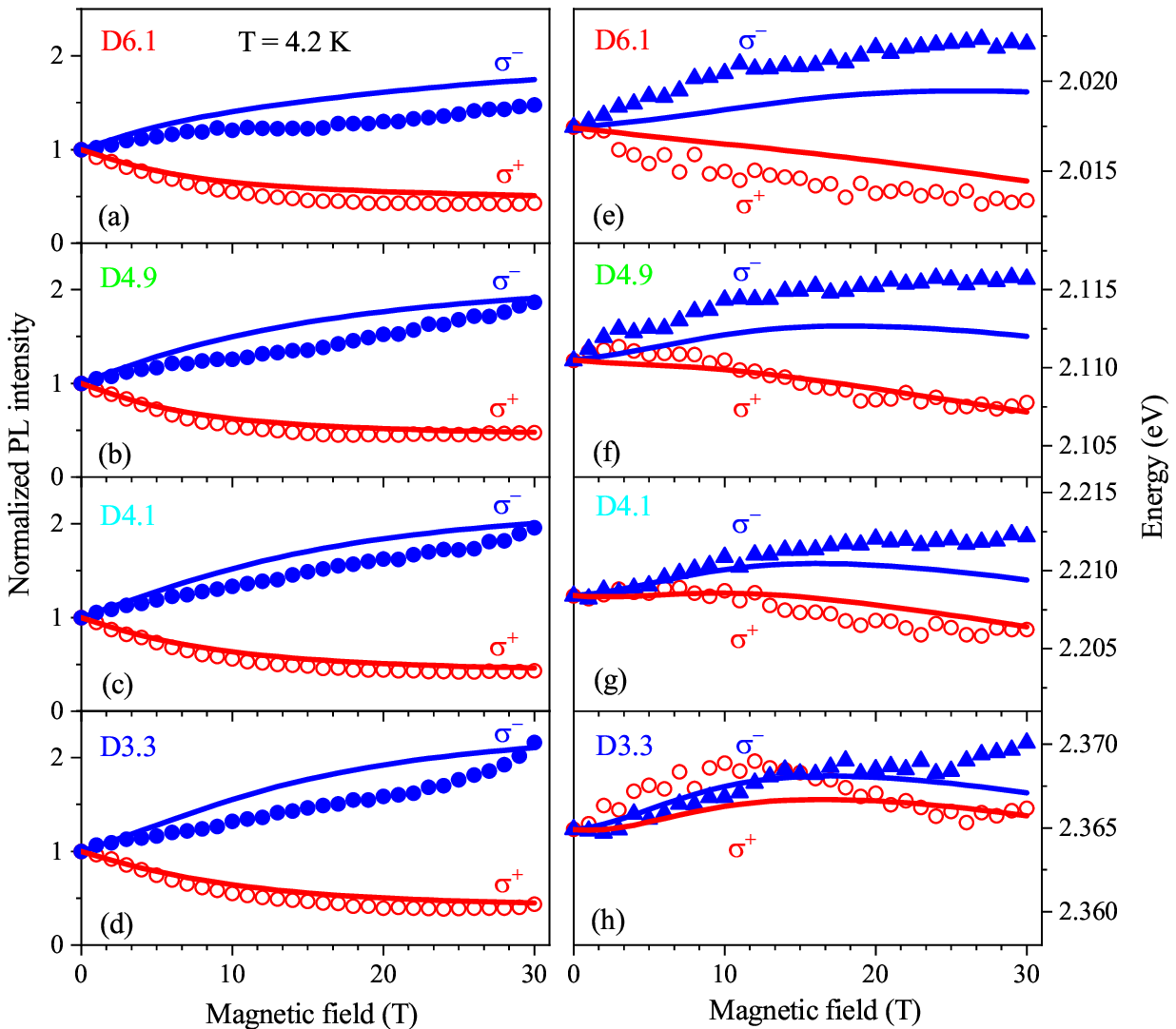}
	\caption{PL intensity and spectral shifts in magnetic field. (a-d) Time-integrated intensity of the ${\sigma}^{+}$ (red) and ${\sigma}^{-}$ (blue) polarized PL as function of the magnetic field in CdSe NCs. (e-h) Magnetic field dependences of the corresponding PL peak energies. For all panels the symbols correspond to the experimental data, the curves show the results of calculations.  The emission of the dark excitons at the ZPL energy via the admixture of the $0^U$ bright exciton is included. The populations of the dark exciton states $\pm2$ are considered to be determined by the relaxation from the $\pm1^L$ states.}	
	\label{fig:figshiftgaAC}
\end{figure*}

\begin{figure*}[h!]
	\centering
	\includegraphics[width= 17.2 cm]{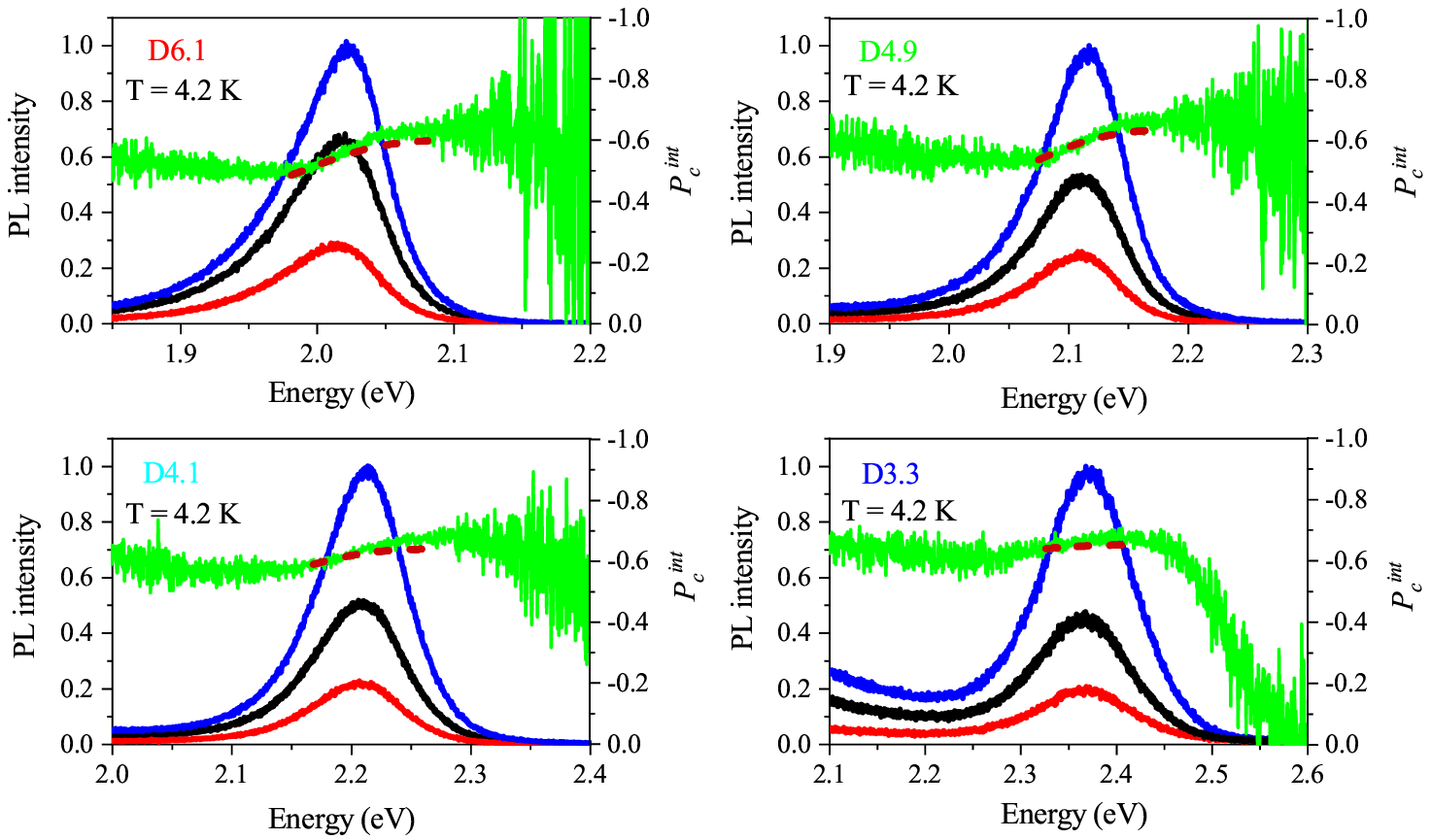}
	\caption{PL spectra of the ${\sigma}^{+}$ (red) and ${\sigma}^{-}$ (blue) polarized components at $B=30$~T and PL spectrum at $B=0$~T (black). Green line shows the experimental spectral dependence of the circular polarization degree at $B=30$~T. Red dash line shows the calculated spectral dependence of the DCP at $B=30$~T. The emission of the dark excitons at the ZPL energy via the admixture of the $0^U$ bright exciton is included. The populations of the dark exciton states $\pm2$ are considered to be determined by the relaxation from the $\pm1^L$ states.}
	\label{fig:spectrDCPgaAC}
\end{figure*}

Considering the acoustic phonon-assisted recombination of the dark excitons in addition to the assumption about the slow spin relaxation between the $\pm2$ states, we perform fits of the experimental data (Figures \ref{fig:figdcpgaac}, \ref{fig:figshiftgaAC}, \ref{fig:spectrDCPgaAC}) with $g_h=-1.2$ for D6.1; $-1.4$ for D4.9; $-1.6$ for D4.1; and $-1.6$ for D3.3. The other fit parameters given in Table~\ref{tab:tablefitAC} remain unchanged, except for $c_{1PL}=0.08$ for the sample D3.3.

\clearpage
\subsubsection*{S4.8 Dark exciton and hole $g$-factors determined from fitting} 

\begin{figure*}[h!]
	\centering
	\includegraphics[width= 17.2 cm]{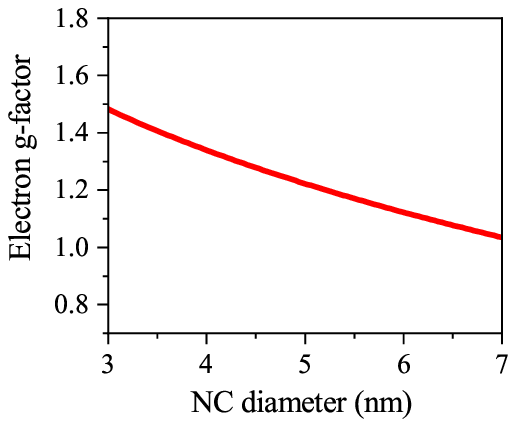}
	\caption{Size dependence of the electron $g$-factor in CdSe NCs according  to Refs.~\citenum{Gupta2002,Tadjine2017,Hu2019}.}
	\label{fig:figge}
\end{figure*}

\begin{figure*}[h!]
	\centering
	\includegraphics[width= 17.2 cm]{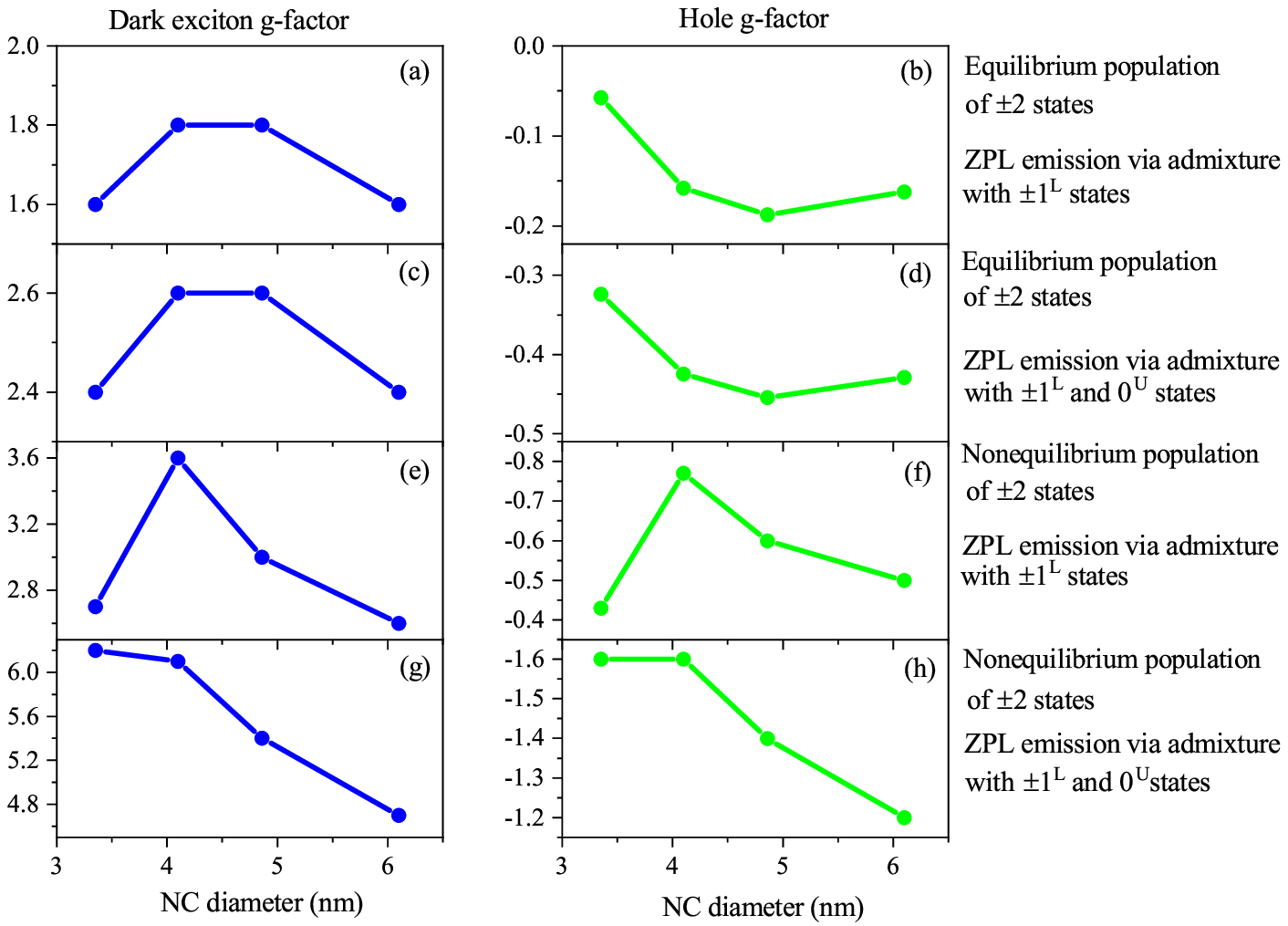}
	\caption{ Size dependence of the dark exciton and hole $g$-factors determined from fits using the following assumptions: (a,b) Fast spin relaxation between the $\pm2$ states and ZPL emission via the $\pm1^L$ admixture; (c,d) Fast spin relaxation between the $\pm2$ states and ZPL emission via the $\pm1^L$ and $0^U$ admixture; (e,f) Slow spin relaxation between the $\pm2$ states and ZPL emission via the $\pm1^L$ admixture; (g,h) Slow spin relaxation between the $\pm2$ states and ZPL emission via the $\pm1^L$ and $0^U$ admixture.
	}
	\label{fig:figgfac}
\end{figure*}

In Figure~\ref{fig:figgfac} we show the dark exciton and hole $g$-factors determined from fitting of the experimental data. These $g$-factors are releated to each other by the equation $g_h=(g_e-g_F)/3$. Here $g_e$ is electron $g$-factor. The size dependence of the electron $g$-factor (see Figure \ref{fig:figge}) is well studied and can be found in Refs.~\onlinecite{Gupta2002,Tadjine2017,Hu2019}. One can see in Figure \ref{fig:figgfac} that the consideration of the linearly polarized ZPL emission and the non-equilibrium population of the dark exciton states significantly modifies the derived dark exciton $g$-factor (from $\approx1.6$ to $\approx$5) and hole $g$-factor (from $\approx-0.1$ to $\approx-1.4$). These results indicate that the value of the hole $g$-factor determined from the analysis of the polarized PL under nonresonant excitation depends on the number of considered recombination channels of the dark exciton and its spin relaxation.  

\hphantom{\cite{Gupta2002,Tadjine2017,Hu2019}}

\end{document}